\documentclass[10pt]{iopart}

\def \BE {\begin{equation}}
\def \EE {\end{equation}}
\def \nn {\nonumber}

\def \bl {\mbox{\boldmath{$\ell$}}}

\def \bn {\mbox{\boldmath{$n$}}}

\def \hbm #1 {\mbox{\boldmath{$\hat m^{(#1)}$}}}
\def \bm {\mbox{\boldmath{$m$}}}

\def \bu  {\mbox{\boldmath{$u$}}} 
\def \bv {\mbox{\boldmath{$v$}}} 
\def \bw {\mbox{\boldmath{$w$}}} 
\def \na {\alpha}
\def \nb {\beta}
\def \ng {\gamma}
\def \nd {\delta}
\def \nf {\phi}
\def \nl {\lambda}
\def \ne {\varepsilon}
\def \tA {\tilde A}

\def \cA {{\cal A}}
\def \cB {{\cal B}}
\def \cC {{\cal C}}
\def \cM {{\cal M}}
\def \cN {{\cal N}}
\def \cO {{\cal O}}
\def \cP {{\cal P}}
\def \cR {{\cal R}}
\def \cS {{\cal S}}
\def \BEAH {\begin{eqnarray*}}
\def \EEAH {\end{eqnarray*}}
\def \BEA {\begin{eqnarray}}
\def \EEA {\end{eqnarray}}
\def \BDM {\begin{displaymath}}
\def \EDM {\end{displaymath}}
\def \Mi {\stackrel{i}{M}}
\def \Mj {\stackrel{j}{M}}
\def \Mk {\stackrel{k}{M}}

\def \Ms {\stackrel{s}{M}}
\def \pul {{{\footnotesize{\frac{1}{2}}}}}
\def \mS {\mbox{\boldmath{$S$}}} 
\def \mA {\mbox{\boldmath{$A$}}}
 \def \mL  {\mbox{\boldmath{$L$}}} 

\def \mPsi {\mathbf \Psi}

\def \T {\bigtriangleup  }
\def \Am {A_{{\rm max}}}
\def \alm {{\alpha}_{{\rm max}}}
\def \S {S}
\def \A {A^2}  

\newcommand{\lp}{\left(}
\newcommand{\rp}{\right)}

\usepackage{latexsym}
\newcommand{\qed}{
\ifmmode \quad\Box
\else
\leavevmode\unskip\penalty9999 \hbox{}\nobreak\hfill
\quad\hbox{$\Box$}
\fi}

\newtheorem{proposition}{Proposition}

\newtheorem{lemma}[proposition]{Lemma}

\newenvironment{proof}{{\em Proof.}}{\qed\par\medskip}

\begin{document}
\title{Bianchi identities in higher dimensions}
\author{V. Pravda\dag, A. Pravdov\' a\dag, A. Coley\ddag, R. Milson\ddag} 

\address{\dag\ Mathematical Institute, 
Academy of Sciences, \v Zitn\' a 25, 115 67 Prague 1, Czech Republic}
\address{\ddag\ Dept. Mathematics and Statistics, Dalhousie U., 
Halifax, Nova Scotia B3H 3J5, Canada}

\eads{\mailto{pravda@math.cas.cz}, \mailto{pravdova@math.cas.cz},
\mailto{aac@mathstat.dal.ca}, \mailto{milson@mathstat.dal.ca}}

\begin{abstract}
A higher dimensional frame formalism is developed in order to study 
implications of the Bianchi identities for the Weyl tensor in vacuum spacetimes 
of the algebraic types III and N in arbitrary dimension $n$. It follows that the principal null  congruence
is geodesic and expands isotropically in  two dimensions and  does not expand in $n-4$ spacelike dimensions or does
not expand at all. It is  shown that the existence of such principal geodesic null congruence 
in vacuum (together with an additional condition on twist) implies an algebraically special spacetime. We also
use the Myers-Perry metric as an explicit example of a vacuum
type D spacetime to show that principal geodesic null congruences in vacuum type D spacetimes do not share this property.
\end{abstract}
\section{Introduction}

For dimensions $n<4$ the Weyl tensor vanishes identically
 and for $n=4$ it has very special properties. 
It is of interest to determine 
which of the properties of four-dimensional (4D) spacetimes can be straightforwardly generalized to higher
dimensions and which need to be modified or do not hold at all. 

Recently a  classification of algebraic tensor types in Lorentzian manifolds of arbitrary dimension
was developed \cite{Algclass}. For the Weyl tensor in 4D this classification  reproduces
the Petrov classification and for the Ricci tensor in 4D  the Segre classification. 

In 4D it follows from the Bianchi identities{\footnote{Note that in this paper we use two
different  operations denoted by $\{\}$. In the first case $\{\}$ acts on three indices
and stands for $R_{ab\{cd;e\}}=R_{abcd;e}+R_{abde;c}+R_{abec;d}$. 
In the other case $\{\}$ acts on four indices and is given by (\ref{zavorka}). }  
}
\BE
R_{ab\{cd;e\}}=0 \label{Bia}
\EE
 that in algebraically special vacuum spacetimes the
multiple principal null direction of the Weyl tensor is geodesic and shearfree.
In this paper a higher dimensional frame formalism is developed in order
to study  implications of the Bianchi identities, which are given in \ref{appbianchi},
for vacuum   spacetimes of algebraic  types N and III in higher dimensions. 
Although in most applications it is necessary
to perform calculations in a given spacetime dimension $n$, 
in this paper we present results
without specifying the dimension and hence these results are valid in any dimension.

In 4D, for algebraically
special vacuum spacetimes some of the tetrad components of the Bianchi identities in the Newman-Penrose formalism
\cite{NPform} lead to simple algebraic equations 
(i.e. equations with no derivatives). In higher dimensions these 
algebraic equations are much more complex and the number of independent equations, as well as the number of unknowns,
depend on the dimension of the spacetime. 
 
We show that in vacuum type III and N spacetimes of arbitrary dimension the multiple principal 
null direction (PND) is geodesic. 
For type N spacetimes (Sec. \ref{typeN}) the symmetric expansion matrix $\mS$ has just one non-vanishing 
doubly degenerate eigenvalue. Consequently, the principal  geodesic null congruence
expands isotropically in  two dimensions and  does not expand in $n-2$ dimensions.
Thus, shear does not vanish for $n>4$. The antisymmetric twist matrix $\mA$ has only one independent 
component. All other components may be set to zero by appropriately choosing the frame.
For type III, we prove similar results in the generic case (Sec. \ref{PIII})
and in the non-twisting case (\ref{appIIInon}). The complete
proof, including all possible degenerate cases, is presented for five-dimensional spacetimes (\ref{app5d}). 

In Section \ref{Goldberg},
we  show that for a vacuum spacetime the properties of $\mS$ and $\mA$ matrices
mentioned above imply that the spacetime is algebraically special. We also show that an
arbitrary vacuum spacetime admitting a non-expanding and non-twisting geodesic null congruence
(i.e., a higher dimensional generalization of the vacuum Kundt class) is algebraically special.

These statements cannot be regarded as a generalization of the Goldberg--Sachs theorem
for higher dimensions (see Sec. \ref{Goldberg} for details) since in \ref{typeD}
it is shown that the Myers--Perry metric, which is of the algebraic type D, has the expansion matrix
$\mS$ with one doubly degenerate eigenvalue and also another non-vanishing 
eigenvalue.

In Section \ref{secdisc}, we conclude with discussion of potential applications.

\section{Preliminaries}

The Newman--Penrose formalism \cite{NPform} in 4D
is based on using a null tetrad $\bl,\bn, \bm$ and $ {\mbox{\boldmath{$\bar m$}}}
$, where $\bl$ and $\bn$ are real null vectors 
and $\bm$ and $ {\mbox{\boldmath{$\bar m$}}}$ are complex null vectors 
  instead of an orthonormal basis, thus taking  advantage of the null cone structure of spacetimes.
For $n$-dimensional calculations it seems to be more practical to choose a pair of null vectors
$\bl,\ \bn$ and an orthonormal set of real spacelike vectors $\bm^{(i)}$. We  thus need two types of
indices: indices $a, b, \dots$ with values $0,\dots,\  n-1$ and indices $i, j,  \dots$ going from 2 to $n-1$.
We will observe Einstein's summation convention for both  of these types of indices. However, for 
indices $i, j, \dots$, there is no difference between covariant and contravariant components
an thus we will not distinguish between subscripts and superscripts.

The frame
\BE
\bm^{(0)}=\bn,\ \bm^{(1)}=\bl,\ \bm^{(i)} \label{tetrad}
\EE
thus satisfies
\BDM
\ell^a \ell_a= n^a n_a =\ell^a m^{(i)}_{a}=n^a m^{(i)}_a= 0,\ \   \ell^a n_a = 1, \ \
m^{(i)a}m^{(j)}_a=\delta_{ij},
\EDM
 and the metric has the form
\BDM
g_{a b} = 2\ell_{(a}n_{b)} + \delta_{ij} m^{(i)}_a m^{(j)}_b. \label{metric} 
\EDM

If one would like to completely generalize the Newman--Penrose formalism for higher
dimensions it would be necessary to denote Ricci rotation-coefficients and all independent components
of the Riemann, Weyl and Ricci tensors and then rewrite all frame components
of the Bianchi identities (\ref{Bia}) and the Ricci identities 
\BDM
V_{a;bc}=V_{a;cb}+R^s_{\ abc} V_s ,
\EDM  
where ${\mbox{\boldmath{$V$}}}$ is an arbitrary vector, as well as expressions for commutators \cite{VSID} of
 covariant derivatives in directions of the frame vectors 
\BE
D \equiv \ell^a \nabla_a, \ \ \ \bigtriangleup  \equiv n^a \nabla_a, \ \ \ \delta_i \equiv m^{(i)a} \nabla_a .
\EE
However, in this paper we are only interested in studying consequences of the Bianchi identities,
which are given in detail in \ref{appbianchi}, and thus we do not introduce the Ricci identities
except for equation (\ref{Riccilmm}) given in  Section \ref{Goldberg}.

\subsection{Decomposition of the Riemann and Weyl tensors}

In order to construct a basis in the space of 4-rank tensors  with symmetries
\BE
R_{abcd}=\frac{1}{2} (R_{[ab] [cd]} + R_{[cd] [ab]}),  \label{Rsym1}
\EE
we introduce the operation \{ \}
\BE
w_{\{a} x_b y_c z_{d\}} \equiv \frac{1}{2}(w_{[a} x_{b]} y_{[c} z_{d]}+ w_{[c} x_{d]} y_{[a} z_{b]}).
\label{zavorka}
\EE

When decomposing the Riemann tensor in terms of the frame vectors we also have to take into account
that
\BE
R_{a\{bcd\}}=0. \label{Rsym2}
\EE
Now let us  decompose the Riemann tensor in its frame components and
sort them 
by their boost weights (see \cite{Algclass}): 
\BEAH
  R_{abcd} = 
  \overbrace{
    4 R_{0i0j}\, n^{}_{\{a} m^{(i)}_{\, b}  n^{}_{c}  m^{(j)}_{\, d\: \}}}^2  \\
  \nonumber
  +\overbrace{
    8R_{010i}\, n^{}_{\{a} \ell^{}_b n^{}_c m^{(i)}_{\, d\: \}} +
    4R_{0ijk}\, n^{}_{\{a} m^{(i)}_{\, b} m^{(j)}_{\, c} m^{(k)}_{\, d\: \}}}^1  
  \nonumber \\
    \begin{array}{l}
      +4 R_{0101}\, \, n^{}_{\{a} \ell^{}_{ b} n^{}_{ c} \ell^{}_{\, d\: \}} 
\;  + \;  4 R_{01ij}\, \, n^{}_{\{a} \ell^{}_{ b} m^{(i)}_{\, c} m^{(j)}_{\, d\: \}}  \\[2mm]
      +8 R_{0i1j}\, \, n^{}_{\{a} m^{(i)}_{\, b} \ell^{}_{c} m^{(j)}_{\, d\: \}}
   +  R_{ijkl}\, \, m^{(i)}_{\{a} m^{(j)}_{\, b} m^{(k)}_{\, c} m^{(l)}_{\, d\: \}}
    \end{array}
\Biggr\}^0
  \label{eq:rscalars}\\ 
   + \overbrace{
    8 R_{101i}\, \ell^{}_{\{a} n^{}_b \ell^{}_c m^{(i)}_{\, d\: \}} +
    4 R_{1ijk}\, \ell^{}_{\{a} m^{(i)}_{\, b} m^{(j)}_{\, c} m^{(k)}_{\, d\: \}}}^{-1} 
  \nonumber \\
  + \overbrace{
      4 R_{1i1j}\, \ell^{}_{\{a} m^{(i)}_{\, b}  \ell^{}_{c}  m^{(j)}_{\, d\: \}}}^{-2} .
\EEAH
The Riemann frame components in this relation are subject to  constraints
following from (\ref{Rsym1}) and (\ref{Rsym2}),
\BEA 
  &&R_{0[i|0|j]} = 0,\nn\\
  \nonumber
  &&R_{0i(jk)}=R_{0\{ijk\}} = 0,\\
  &&R_{ijkl} = R_{\{ijkl\}},\quad R_{i\{jkl\}}=0,\quad 
  R_{01ij}= 2 R_{0[i|1|j]}, \label{eq:rcomps}\\
  \nonumber
  &&R_{1i(jk)}=R_{1\{ijk\}} = 0, \\
  \nonumber
  &&R_{1[i|1|j]} = 0.
\EEA
Let us check that we have an appropriate number of independent
frame components.
It is well known that an $n$-dimensional Riemann tensor
has
\begin{equation}
  \label{eq:rdim}
  \frac{n^2(n^2-1)}{12}
\end{equation}
independent components. Counting independent frame components of various boost weights,
we obtain
\BDM
\fl
  2\overbrace{\lp \frac{m(m+1)}{2}\rp}^{2,-2} +
  2\overbrace{\lp \frac{(m+1)m(m-1)}{3}+m\rp}^{1,-1}+ 
  \overbrace{\frac{m^2(m^2-1)}{12}+ m^2+1}^0 ,
\EDM
where $m=n-2$. This is in agreement with (\ref{eq:rdim}).

Similarly, it is possible to decompose the Weyl tensor  but due to its tracelessness
we have the following additional conditions:
\BEA
  \label{eq:rfree}
  && C_{0i0i} =  C_{1i1i} =0,\\
  \nonumber
  &&C_{010i} = C_{0jij},\quad C_{101i} = C_{1jij},\\
  \nonumber
   && 2C_{0i1j}=C_{01ij}- C_{ikjk},\quad 
  C_{0101} = -\pul C_{ijij}.
\EEA 
It is well known that an $n$-dimensional  Weyl tensor
has
\begin{equation}
  \label{eq:wdim}
  \frac{(n+2)(n+1)n(n-3)}{12}  
\end{equation}
independent components.  By
counting the independent scalars of various boost weights,
we obtain
\BDM
\fl
\overbrace{2 \lp \frac{(m+2)(m-1)}{2} \rp}^{2,-2} + \overbrace{2 \lp \frac{(m+1)m(m-1)}{3} \rp}^{1,-1} 
+\overbrace{\frac{m^2(m^2-1)}{12} + \frac{m(m-1)}{2}}^0 ,
\EDM
which is in agreement with (\ref{eq:wdim}).

The primary algebraic classification of the Weyl tensor
in higher dimensions  \cite{Algclass} is based on  whether all Weyl frame components
of a  boost weight higher then a specified number can be transformed away
by an appropriate choice of the null direction $\bl$. If it is indeed possible, we call the corresponding
direction a Weyl aligned null direction (WAND) of an appropriate order (0, 1, 2, 3). If the set of WANDs of a given order 
is discrete, we call the corresponding directions principal null directions (PND).

Type III and N spacetimes admitting a WAND of order 2 and 3, respectively, were introduced in \cite{PP6}.
We will use the notation based on that given in \cite{PP6}
which is suitable for these algebraic classes but becomes rather cumbersome in more general cases.  
Let us in accordance with \cite{Algclass,PP6} state that a spacetime is of the algebraic type III if there exists a 
frame (\ref{tetrad}) in which the Weyl tensor has the form
\BE 
\fl
C_{abcd} = 8 \Psi_{i} \, \ell^{}_{\{a} n^{}_{ b} \ell^{}_{ c} m^{(i)}_{\, d\: \}}+
8 \Psi_{ijk} \, m^{(i)}_{\{a} m^{(j)}_{\: b} \ell^{}_{ c} m^{(k)}_{\, d\: \}}+ 
8 \Psi_{ij} \, \ell^{}_{\{a} m^{(i)}_{\: b} \ell^{}_{ c} m^{(j)}_{\, d\: \}} \label{WeylIIIN},
\EE
with $\Psi_{ijk} \not= 0$. The case with
$\Psi_{ijk} = 0$ (and consequently from (\ref{eqPsi_i}) also $\Psi_{i} = 0$) is of the algebraic type N.
Note, however, that in this paper the operation $\{ \}$ differs by a factor $1/8$ from that given in \cite{PP6}.

 The  components of the Weyl tensor $\Psi_{i}, \ \Psi_{ijk}$  and $\Psi_{ij}$ are given by
\BDM
\Psi_{i} = C_{101i}, \ \  \Psi_{ijk}= \frac{1}{2} C_{1kij}, \ \ \Psi_{ij} = \frac{1}{2} C_{1i1j} .
\EDM
Note that $\Psi_{ij}$ is symmetric and traceless. $\Psi_{ijk}$ is antisymmetric in the first two indices
and from (\ref{eq:rcomps}) and (\ref{eq:rfree}) it also follows that
\BEA
\Psi_i=2 \Psi_{ijj}, \label{eqPsi_i}\\ 
   \Psi_{\{ijk\}}=0. \label{eqPsicykl}
\EEA

\subsection{Decomposition of covariant derivatives of the frame vectors}

Let us denote components of covariant derivatives of the frame 
vectors $\bl,\bn, \bm^{(i)}$ by $L_{ab},N_{ab}$ and $\Mi_{ab}$, respectively,
\BDM
\fl 
\ell_{a;b}=L_{cd} m^{(c)}_a m^{(d)}_{b} \ , \ \ n_{a;b}=N_{cd} m^{(c)}_a m^{(d)}_{b} \ , \ \
m^{(i)}_{a;b}=\Mi_{cd} m^{(c)}_a  m^{(d)}_{b}  \ .
\EDM
Since the norm of  all frame vectors is constant,  it follows that
\BDM
L_{0a}=N_{1a}=\Mi_{ia}=0.
\EDM
Also from the fact that all scalar products of the frame vectors are constant, we get
\BE
\!\!\!\!\!\!\!\!\!\!\!\!\!\!\!\!\!\!
N_{0a}+L_{1a}=0, \quad \Mi_{0a} + L_{ia} = 0, \quad \Mi_{1a}+N_{ia}=0,
\quad \Mi_{ja}+\Mj_{ia}=0. \label{const-scalar-prod}
\EE  
We thus arrive at
\BEA
\fl
\ell_{a ; b} &=& L_{11} \ell_a \ell_b + L_{10} \ell_a n_b + L_{1i} \ell_a m^{(i)}_{\, b}  +
L_{i1} m^{(i)}_a \ell_b   + L_{i0} m^{(i)}_{\, a} n_{b} + L_{ij} m^{(i)}_{\, a} m^{(j)}_{\, b}  , \label{dl} \\
\fl
n_{a ; b  } &=&\! -\! L_{11} n_a \ell_b -\! L_{10} n_{a } n_{b }-\! L_{1i} n_a m^{(i)}_{\, b}  +
N_{i1} m^{(i)}_{\, a} \ell_b  + N_{i0} m^{(i)}_{\, a} n_b  + N_{ij} m^{(i)}_{\, a} m^{(j)}_{\, b}  , \label{dn} \\
\fl
m^{(i)}_{a ; b } &=&\! -\! N_{i1} \ell_a \ell_b -N_{i0} \ell_a n_b -{L}_{i1} n_a \ell_b  
-L_{i0} n_a n_b -N_{ij} \ell_a m^{(j)}_{\, b}   \nonumber \\
\fl 
&&\! +\! {\Mi}_{j1} m^{(j)}_{\, a} \ell_b  -L_{ij} n_a m^{(j)}_{\, b} 
+ {\Mi}_{j0} m^{(j)}_{\, a} n_b + {\Mi}_{kl} m^{(k)}_{\, a} m^{(l)}_{\, b} . \label{dm} 
\EEA

\subsection{Null geodesic congruences}

In Sections \ref{typeN} and \ref{PIII}, we will show that the multiple PND in type III and N vacuum spacetimes is geodesic. 
Let us  thus study properties of null geodesic congruences in higher dimensions. Analogous but in some
cases not fully equivalent definitions are given in the Appendix of the paper \cite{Frolov}.

The congruence corresponding to $\bl$ is geodesic if 
\BDM
\ell_{a ; b } \ell^{b} \propto \ell_{a } \   \Leftrightarrow \  L_{i0} = 0.
\EDM
It is always possible to rescale
$\bl$ (and consequently also $\bn$) in such a way that 
$\ell_{a ; b } \ell^{b} = 0$ and thus also $L_{10}=0$.
From now on we will use this parametrization.
Then the covariant derivative of the vector $\bl$ is
\BE
\ell_{a ; b } = L_{11} \ell_a \ell_b +L_{1i} \ell_a m^{(i)}_b  +
L_{i1} m^{(i)}_a \ell_b  + L_{ij} m^{(i)}_{a} m^{(j)}_{b} \label{dlgeod} .
\EE
Let us decompose ${\mL}$ into its symmetric and antisymmetric parts, ${\mS}$ and
${\mA}$,
\BDM
L_{ij}=S_{ij} + A_{ij}, \quad S_{ij} = S_{ji} , \ A_{ij}= -A_{ji}   ,
\EDM
where
\BDM
S_{ij}=\ell_{(a ; b)} m^{(i)a} m^{(j)b} \ , \ \ \ A_{ij}=\ell_{[a ; b]} m^{(i)a} m^{(j)b}.      
\EDM
We define the expansion $\theta$ and the shear matrix $\sigma_{ij}$ as follows:
\BEA 
\theta &\equiv &\frac{1}{n-2} \ell^{a}_{\ ; a} = \frac{1}{n-2}[\mS] ,\\
\sigma_{ij} &\equiv& \left( 
\ell_{(a ; b)} 
- \theta \delta_{kl}   m^{(k)}_a m^{(l)}_b \right)
m^{(i)a} m^{(j)b}  
=S_{ij}-\frac{[\mS]}{n-2}\delta_{ij}. \label{shearmatrix}
\EEA
We will also denote $\sigma^2\equiv\sigma_{ij} \sigma_{ij}$.
Let us, for simplicity,  call $\mA$ the {\it twist matrix} and $\mS$  the {\it expansion matrix}, though 
$\mS$ contains information about both, expansion and shear.
For simplicity we also introduce quantities
\BE
\S\equiv\pul[\mS],\ \ \A\equiv\pul A_{ij}A_{ij}.
\EE

\section{Type N vacuum spacetimes}
\label{typeN}

For type N spacetimes the Weyl tensor (\ref{WeylIIIN}) has the form 
\BE
C_{abcd} = 8 \Psi_{ij} \, \ell^{}_{\{a} m^{(i)}_{\, b} \ell^{}_{ c} m^{(j)}_{\, d\: \}}, \label{WeylN}
\EE
where $\mPsi$ is symmetric and traceless.
Bianchi equations, given in \ref{appbianchi}, now reduce to 
\BEA
D \Psi_{ij}&=&-2 \Psi_{k(i} \Mk_{j)0}- \Psi_{ik} L_{kj} -2 \Psi_{ij} L_{10} ,\label{BiaN1} \\
\delta_{[k} \Psi_{j]i} &=&-  \Psi_{is} \Ms_{[jk]} + \Ms_{i[j}\Psi_{k]s} 
+ \Psi_{i[j} L_{k]1} + 2 L_{1[j} \Psi_{k]i} ,\label{BiaN2} \\
\ \ \ \ \ \ 0&=& L_{k[i} \Psi_{j]k},\label{BiaN3} \\
\ \ \ \ \ \ 0&=&  \Psi_{i[j}L_{k]0}, \label{BiaN4} \\
\ \ \ \ \ \ 0&=&  L_{k[j} \Psi_{m]i} + L_{i[m} \Psi_{j]k}, \label{BiaN5} \\
\ \ \ \ \  \ 0&=&  \Psi_{i\{k}A_{jm\}}. \label{eqBiaA}
\EEA

Let us  first
show that from equation (\ref{BiaN4}) it follows that 
the multiple PND $\bl$ in type N  vacuum spacetimes is geodesic.
For simplicity we will  denote $L_{i0}$ as $L_i$.

The contraction of $i$ with $k$ in (\ref{BiaN4})  leads to
\BDM
\Psi_{ij} L_{i} = 0 .
\EDM
Now the contraction of (\ref{BiaN4})  with $\Psi_{ik}$ gives
\BDM
\Psi_{ik} \Psi_{ik} L_{j} = 0.
\EDM
Note that in type N spacetimes $\Psi_{ik} \Psi_{ik}  > 0$ and thus the previous equation 
implies $L_{i} =0$ and  $\bl$ is indeed geodesic.

By substituting $\mL=\mA+\mS$ into (\ref{BiaN5}) and using (\ref{eqBiaA}), we obtain
\BE
S_{k[j} \Psi_{m]i} + S_{i[m} \Psi_{j]k} = 0.  \label{eqBiS}
\EE

Let us now study in detail consequences of  equations (\ref{BiaN3}), (\ref{eqBiaA}) and (\ref{eqBiS}).
It is possible to find a general solution of these equations by taking
into account their various contractions (hereafter we assume $\mS\not=0$ and $\mA\not=0$; 
the non-twisting case can be obtained as a special case with $\mA=0$).

Contracting $i$ with $j$ in (\ref{eqBiaA}) gives
\BE
\Psi_{ik} A_{im} = \Psi_{im} A_{ik}\ \  \Rightarrow \ \ 
\mPsi \cdot \mA + \mA \cdot \mPsi  = 0  \label{anticomA}
\EE
and contracting $k$ with $j$ in (\ref{eqBiS}) leads to
\BE
\Psi_{ij} S_{jm}
+\Psi_{jm} S_{ij} =  
2 \S 
\Psi_{im}  \ \ \Rightarrow\ \  \mPsi \cdot \mS + \mS \cdot \mPsi = 2\S
\mPsi  .
\label{anticomS}
\EE
Previous two equations imply
\BDM
\mPsi \cdot \mL + \mL \cdot \mPsi = 2\S 
\mPsi .
\EDM
By contracting $i$ with $m$ in (\ref{anticomS}) we get
\BE
\Psi_{ij} S_{ij} = 0\ \  \Rightarrow\ \   \Psi_{ij} L_{ij} = 0 .  \label{conPS}
\EE
From (\ref{BiaN3}) and (\ref{anticomA}) it follows that
\BE
\Psi_{ki} S_{kj} = \Psi_{kj} S_{ki} \ \ \Rightarrow\ \  \mPsi \cdot \mS - \mS \cdot \mPsi = 0. \label{comS}
\EE
Let us denote a trace $[\mPsi \cdot \mPsi]$ as $p$. Now we are in a position to formulate the following
lemma.

\begin{lemma}
\label{lem}
In vacuum type N spacetimes with $\mL \not=0$ following relations are satisfied:
\BEA
(a)\ \ && \mPsi \cdot \mS = \S  
\mPsi,\\
(b)\ \ && \mS \cdot \mS = \S 
\mS ,\\
(c)\ \ && \mA \cdot \mS = \mS \cdot \mA = \S 
 \mA, \\
(d)\ \ && \S 
\mA \cdot \mA = -\A 
\mS ,\\
(e)\ \ && 2\S 
\mPsi \cdot \mPsi = p \mS  ,\\
(f)\ \ && \S 
 \mL \cdot \mL^T = \S 
\mL^T \cdot \mL = (\S^2 
+\A
) \mS  .
\EEA
\end{lemma}
\begin{proof}
 {\bf (a)} This equation is a direct consequence of (\ref{anticomS}) and (\ref{comS}). \\[2mm]
{\bf (b)}
By multiplying (\ref{eqBiS}) by $S_{ip}$ and using Lemma 1(a) 
we get
\BDM
\Psi_{kj} S_{im} S_{ip}+\S 
\Psi_{mp} S_{kj}=\Psi_{km} S_{ij} S_{ip}+\S 
\Psi_{jp} S_{km} .  \label{eqBiaS}
\EDM
An appropriate linear combination of this equation with (\ref{eqBiS}) leads to
\BDM
\Psi_{kj} (S_{im} S_{ip} - \S 
S_{mp}) = \Psi_{km} (S_{ij} S_{ip} -\S 
S_{pj}),
\EDM
which, denoting
$
X_{mp} \equiv S_{im} S_{ip} - \S 
S_{mp},
$
takes the form
\BE
\Psi_{kj} X_{mp} = \Psi_{km} X_{jp}  . \label{eqPsiX}
\EE
Note that the matrix {\mbox{\boldmath{$X$}}}  is symmetric.
Now contracting $k$ with $j$ gives 
\BE
\Psi_{jm} X_{jp} = 0  .\label{prodPsiX}
\EE
Multiplying  (\ref{eqPsiX}) with $X_{mr}$ and using (\ref{prodPsiX})
leads to 
$
X_{mp} X_{mr} =0,
$
which, contracting $p$ with $r$, gives
$
X_{mp} X_{mp} =0 
$
and consequently $ X_{mp}=0$.\\
{\bf (c)}
Multiplying (\ref{eqBiaA}) with $S_{kp}$ and using Lemma 1(a)
leads to
\BE
\psi_{ij} A_{km} S_{kp} + \S 
\psi_{ip} A_{mj} + \psi_{im} A_{jk} S_{kp} =0.
\label{eqlemc}
\EE
Now contracting $i$ with $m$ and using (\ref{anticomA}) gives
\BE
\psi_{ij} Y_{ip} = 0 , \label{prodPsiY}
\EE
where
$
Y_{ip} \equiv A_{ki} S_{kp} + \S 
A_{ip}  .
$
Now substituting  $A_{ki} S_{kp}= Y_{ip} -\S 
A_{ip} \ $ into
(\ref{eqlemc}) and using (\ref{eqBiaA}) leads to
\BE
\psi_{ij} Y_{mp} = \psi_{im} Y_{jp}. \label{eqPsiY}
\EE
Multiplying (\ref{eqPsiY}) with $Y_{mr}$ using (\ref{prodPsiY})
gives
$
Y_{mp} Y_{mr} =0 $ and consequently  $ Y_{mp} =0$.\\
{\bf (d)}
Let us define another symmetric matrix {\mbox{\boldmath{$B$}}}  by
$
B_{ik} \equiv A_{ij} A_{jk} .
$
Note that
$
[{\mbox{\boldmath{$B$}}} ]=-2 \A. 
$
Multiplying (\ref{anticomA}) by $A_{mr}$ and again applying (\ref{anticomA}) leads
to
\BE
\psi_{ik} B_{ir} = \psi_{ir} B_{ik}. \label{eqprodPsiB}
\EE
Multiplying (\ref{eqBiaA}) by $A_{mr}$ leads to
\BE
\psi_{ij} B_{kr} - \psi_{ik} B_{jr} + \psi_{im} A_{jk} A_{mr}=0 , \label{eqPsiB}
\EE
which after contracting $k$ with $r$ gives
\BE
\psi_{ik} B_{jk} = -\A 
\Psi_{ij} .\label{eqpsiBrov}
\EE
Multiplying (\ref{eqPsiB}) with $B_{kr}$ and using (\ref{eqprodPsiB}), (\ref{eqpsiBrov}) and
(\ref{anticomA}) leads to
$
B_{kr} B_{kr} = 2 {A}^4 
$
and similarly multiplying (\ref{eqPsiB}) by $S_{kr}$ results in
$
B_{kr} S_{kr} = -2\S 
\A 
$.
Let us now define a symmetric matrix  {\mbox{\boldmath{$Q$}}} by
\BDM
Q_{ij} \equiv \S 
B_{ij} + \A 
S_{ij}.
\EDM
Using previous formulae it turns out that
$
Q_{ij} Q_{ij} = 0 
$ and thus $Q_{ij}=0$.\\
{\bf (e)}
This follows from multiplying (\ref{eqBiS})  by $\Psi_{kj}$.\\[2mm]
{\bf (f)}
This follows directly from Lemma 1(b)--(d).  
\end{proof}

Let us now use  Lemma 1 to prove following lemmas for vacuum type N spacetimes with $\mL \not= 0$:
\begin{lemma}
The matrix $\mS$ has at most two  eigenvalues $\lambda=0$ and $\lambda=\S $.
\end{lemma} 
\begin{proof}
Let us denote the eigenvector of $\mS$ by $\xi$. We thus have
$
S_{ij} \xi_j = \lambda 
\xi_i
$.
By multiplying Lemma 1(b) by $\xi$
we obtain
\BDM
S_{ij} S_{jk} \xi_k = \S 
S_{ik} \xi_k \ \ \Rightarrow\ \  \lambda^2 =\S 
\lambda 
\EDM
and thus $\lambda=0$ or $\lambda= \S 
$.
\end{proof}
\begin{lemma}
 The following statements are equivalent:\\
 (a) Vector $\xi$ is an eigenvector of $\mS$ for the eigenvalue $\lambda=0$.\\
 (b) Vector $\xi$ is an eigenvector of $\mA$ for the eigenvalue $\lambda=0$.\\
 (c) Vector $\xi$ is an eigenvector of $\mPsi$ for the eigenvalue $\lambda=0$.
\end{lemma}
\begin{proof}
  (a) $\Rightarrow$ (b): \  
  Suppose that $ S_{ij} \xi_j = 0$. Then by multiplying Lemma 1(c) in the form 
  $S_{ij} A_{jk} = \S 
A_{ik}$ by $\xi_i$ we obtain
  $
  A_{ik} \xi_i = 0
  $.\\
   (b) $\Rightarrow$ (a): \
  Suppose that $A_{ij} \xi_j = 0$. Then by multiplying Lemma 1(d) in the form 
  $\S 
A_{ij} A_{jk} = -\A 
S_{ik}$ by $\xi_k$ we obtain
  $
  S_{ik} \xi_k = 0
  $.\\
   (a) $\Rightarrow$ (c): \
  Suppose that $S_{ij} \xi_j = 0$. Then by multiplying Lemma 1(a) in the form 
  $ \Psi_{ij} S_{jk} = \S 
\Psi_{ik}$ by $\xi_k$ we obtain
  $
  \Psi_{ik} \xi_k = 0
  $.\\
  (c) $\Rightarrow$ (a): \
  Suppose that $\Psi_{ij} \xi_j = 0$. Then by multiplying Lemma 1(e) in the form 
  $2\S 
\Psi_{ij} \Psi_{jk} = p S_{ik}$ by $\xi_k$ we obtain
 $
  S_{ik} \xi_k = 0
  $.
\end{proof}
\begin{lemma}
Every eigenvector of $\mPsi$ is also an eigenvector of $\mS$.
\end{lemma}
\begin{proof}
The case $\lambda=0$ is solved in Lemma 3. Let us now suppose that
$\Psi_{ij} \xi_j = \lambda \xi_i$, $\lambda \not= 0$.
Then by multiplying Lemma 1(a) in the form 
  $\Psi_{ij} S_{jk} = \S 
\Psi_{ik}$ by $\xi_i$ we obtain
  $
  S_{jk} \xi_j = \S 
\xi_k 
  $.
\end{proof}
\begin{lemma}
  The only possible eigenvalues of $\mPsi$ are $\lambda=0$ and $\lambda=\pm \sqrt{p/2}.$
\end{lemma}
\begin{proof}
Let us now suppose that
$\Psi_{ij} \xi_j = \lambda \xi_i$, $\lambda \not= 0$. Than thanks to Lemma 4 
$S_{ij} \xi_j = \S 
\xi_i$. 
By multiplying Lemma 1(e) in the form $2\S 
\Psi_{ij} \Psi_{jk} = p S_{ik}$ by $\xi_k$ we obtain 
$\lambda^2=p/2$.
\end{proof}

For every symmetric matrix {\mbox{\boldmath{$W$}}}  there exists an orthonormal 
basis of its eigenvectors $\xi^a_i$ in which 
\BDM
W_{ij} = \sum_a \lambda_a  \xi^a_i \xi^a_j.
\EDM
Let us denote eigenvectors of $\mPsi$ corresponding to $\lambda=0$, $\lambda=\sqrt{p/2}$ and $\lambda=-\sqrt{p/2}$
by  $ \bu^{\alpha}$, $\bv^A$  and  $\bw^{\tilde A}$, respectively. Here indices 
${\alpha}, \beta,  \dots$, $A, B, \dots$ and
$\tilde A, \tilde B, \dots $ 
distinguish between vectors $\bu^1, \dots,\ \bu^{\alm}$, $\bv^1, \dots,\ \bv^{\Am}$ 
and $\bw^1, \dots,\ \bw^{\tilde \Am}$.
Now we have 
\BE
\Psi_{ij} = \sqrt{\frac{p}{2}}  (v^A_i v^A_j - w^{\tA}_i w^{\tA}_j). \label{decompsumPsi}
\EE
Note that for indices ${\alpha}, \beta, \dots$, $A, B, \dots$ and
$\tilde A, \tilde B, \dots $ 
 we also observe Einstein's summation convention. 
Thanks to Lemmas 3 and 4 
\BE
S_{ij} = \S 
(v^A_i v^A_j + w^{\tA}_i w^{\tA}_j).  \label{decompsumS}
\EE
Let us now multiply (\ref{eqBiS}) by $v^A_k v^B_j$. This leads to 
\BDM
\delta^{AB} (\sqrt{\frac{p}{2}} S_{il} + \S 
\Psi_{il}) = \S 
\sqrt{2p} \ v^A_l v^B_i
\EDM
 which for $A \not=B$ gives
$
v^A_l v^B_i = 0
$
and thus just one value of index $A$ is possible and consequently $\Am=1$. Let us thus denote $\bv^1$ by $\bv$. 
Similarly, we 
can find that $\tilde \Am=1$ and that  there is thus just one vector $\bw=\bw^1$.  
Equations (\ref{decompsumPsi}) and  (\ref{decompsumS})
now take form
\BE
\Psi_{ij} = \sqrt{\frac{p}{2}}  (v_i v_j - w_i w_j)  , \ \ \ 
S_{ij} = \S 
(v_i v_j + w_i w_j).  \label{decompPsiS}
\EE
Let us now introduce a new vector  {\mbox{\boldmath{$V$}}}  by
\BDM
V_i = A_{ij} v_j.
\EDM
Thanks to Lemma 1(c) and (d) we have
\BEAH
S_{ij} V_i = S_{ij} A_{ik} v_k =\S 
V_j, \quad
V_i v_i = 0, \quad
V_i V_i =  \A 
\EEAH
and thus {\mbox{\boldmath{$V$}}}  is just a multiple of $\bw$ $(w_i=\pm \frac{1}{{A} 
} V_i)$
and we  have freedom to set
$
w_i= \frac{1}{{A} } V_i.
$
Then we have 
\BDM
A_{ij} v_j = {A} 
 w_i  , \ \ A_{ij} w_j = -{A}
v_i .
\EDM
Now by multiplying equation (\ref{eqBiaA}) by $v_i v_j$
we get
\BE
A_{kl} = {A}
 (w_k v_l  - v_k w_l) \label{decompA} .
\EE
Note that $\mPsi, \mS$ and $\mA$ given by (\ref{decompPsiS}) and (\ref{decompA}) satisfy  equations
(\ref{eqBiaA}), (\ref{eqBiS}) and thus represent the general solution of these equations for $\mL \not= 0.$

\section{Type III vacuum spacetimes}
\label{PIII}

For  type III vacuum spacetimes
the Weyl tensor is given by (\ref{WeylIIIN}), where $\Psi_{ijk}$  satisfies 
(\ref{eqPsi_i}) and (\ref{eqPsicykl}). The algebraic equations 
\BEA
 \Psi_{ijk} L_{k} = L_{[i} \Psi_{j]} ,  \label{eqgeoda} \\
\Psi_{kl[j} L_{i]} = \Psi_{ij[k} L_{l]} , \label{eqgeodb} \\
 L_{i[j} \Psi_{k]} + 2 L_{s[j|} \Psi_{si|k]} =0 , \label{Bia33}\\
2 A_{\{ij|} \Psi_{kl|m\}} + L_{l\{i} \Psi_{jm\}k} - L_{k\{i} \Psi_{jm\}l}=0, \label{Bia35} 
\EEA
where we denote $L_{i0}$ as $L_i$,
follow from the Bianchi equations given in \ref{appbianchi}.

Let us show that from (\ref{eqgeoda}) and (\ref{eqgeodb}) it follows 
that the multiple PND $\bl$ in vacuum type  III spacetimes is geodesic.
By contracting $l$ with $j$ in (\ref{eqgeodb}) and using (\ref{eqgeoda}), (\ref{eqPsi_i}),
and  (\ref{eqPsicykl}) we obtain 
\BE
2 \Psi_{ijk} L_j = \Psi_k L_i  . \label{eqgeodc}
\EE
One can see that for the class with $\Psi_k\not=0$ it implies
(as we can see by contracting this formula with $L_i$) that $L_i=0$. 
Similarly, for the case $\Psi_k=0$ these equations also imply that $L_i=0$ 
 (contract (\ref{eqgeodb}) with $L_i$ and employ (\ref{eqgeoda}) and (\ref{eqgeodc})).
Thus, the multiple PND $\bl$ in vacuum type III spacetimes is  geodesic.

%
Now the task is to solve (\ref{Bia33}) and (\ref{Bia35}). This turns out to be quite
complicated. We will be able to find a general solution in an arbitrary dimension 
in the non-twisting case with $A_{ij}$ = 0 (see \ref{appIIInon}). 
We also present  a solution for the twisting case, but we can prove that
it is a general solution only if we assume that $\Psi_{ijk}$ is in a `general form'.
In  \ref{app5d}, we show that this solution
is indeed general in all degenerate cases in five dimensions.  

Let us start with extracting some information from (\ref{Bia33}) and (\ref{Bia35})
by various contractions.
By contracting (\ref{Bia35}) and using (\ref{Bia33}) we get 
\BE\
 L \Psi_{ijk} + 2 L_{[i|s} \Psi_{sk|j]} - 2 S_{sk} \Psi_{ijs} +  L_{[i|k} \Psi_{|j]} =0,
\label{Bia335}
\EE
where $L=2\S$ is the trace of $\mL$.
After adding (\ref{Bia33}) (where we replace indices $i,j,k$ by $k,i,j$, respectively) to (\ref{Bia335}) we obtain
\BE
 L \Psi_{ijk} + 4 S_{[i|s} \Psi_{sk|j]}   - 2 S_{sk} \Psi_{ijs} +  2 S_{[i|k} \Psi_{|j]}   =0 ,  \label{BS1}
\EE
which does not contain $\mA$ -- the antisymmetric part of $\mL$, and similarly 
\BE
L \Psi_{ijk} - 4 \Psi_{sk[i} A_{j]s} - 2 \Psi_{[i} A_{j]k} - 2 \Psi_{ijs} S_{sk} = 0 . \label{eqPsiA}
\EE
Contraction of $i$ with $j$ in (\ref{Bia33}) leads to
\BE
L \Psi_k - 2 L_{ik} \Psi_i + 2 L_{si} \Psi_{sik} = 0 \label{eqLPsi1}
\EE
and contraction of $k$ with $j$ in (\ref{BS1}) (using (\ref{eqPsi_i}) and (\ref{eqPsicykl})) to
\BE
L_{ij} \Psi_j = 2 L_{js} \Psi_{ijs} \label{eqLPsij}
\EE
and (\ref{eqLPsi1}), using (\ref{eqPsicykl}), with (\ref{eqLPsij}) imply
\BE
L\Psi_i-L_{ij}\Psi_j-2L_{ji}\Psi_j+2L_{kj}\Psi_{ijk}=0.\label{eqLPsi2}
\EE
By substituting ${\mL}={\mS}+{\mA}$ into (\ref{eqLPsi1}) and (\ref{eqLPsi2}) we get
\BEA
&&L\Psi_i-2S_{ij}\Psi_j-2A_{ji}\Psi_j-4A_{jk}\Psi_{ijk}=0,\label{preAS1}\\
&&L\Psi_i-3S_{ij}\Psi_j-A_{ji}\Psi_j+2(S_{jk}-A_{jk})\Psi_{ijk}=0\label{preAS2}
\EEA
and their linear combinations 
give
\BEA
&&L\Psi_i+4A_{ij}\Psi_j-4(S_{jk}+2A_{jk})\Psi_{ijk}=0,\label{AS2}\\
&&L\Psi_i-4S_{ij}\Psi_j+4S_{jk}\Psi_{ijk}=0.\label{AS1}
\EEA
By multiplying (\ref{Bia33})  by $L_{ij}$ we arrive to
\BE
L_{ij} L_{ik} \Psi_j = \pul {\ell} \Psi_k , \label{eqell}
\EE 
where ${\ell} = L_{ij} L_{ij}$.

Inspired by the type  N 
case, we again choose an orthonormal basis of eigenvectors of $\mS$. We will
denote vectors corresponding to non-zero eigenvalues of $\mS$ as $\bv^1,\bv^2, \dots, \bv^{\Am}$ and vectors
corresponding to the eigenvalue 0  as $\bu^1,\bu^2, \dots, \bu^{\alm}$. We thus have
\BE
S_{ij} = \sum_{A=1}^{\Am} \lambda_A v_i^A v_j^A. \label{decompIIISa}
\EE
Unfortunately, there are three indices $A$ in this formula and we thus will not use Einstein's summation
convention for indices $A,B, \dots$ (though we will still use it for indices $\alpha, \beta, \dots$).

Let us now decompose other quantities appearing in (\ref{Bia33}) and (\ref{Bia35})
\BEA
\fl \Psi_i = \sum_A a_{A} v_i^A + b_{\alpha } u_i^{\alpha } , \label{ALB}\\
\fl A_{ij} = \sum_{AB} \cA_{AB} (v_i^A v_j^B - v_j^A v_i^B) +
\sum_A \cB_{A\beta } (v_i^A u_j^{\beta } - v_j^A u_i^{\beta }) +
\cC_{\alpha \beta } (u_i^{\alpha } u_j^{\beta } - u_j^{\alpha } u_i^{\beta }) , \nonumber \\
\fl \Psi_{ijk} = \sum_{ABC} \cM_{ABC} (v_i^A v_j^B - v_j^A v_i^B) v_k^C + \sum_{AB} 
\cN_{AB\gamma } (v_i^A v_j^B - v_j^A v_i^B) u_k^{\gamma} 
\nonumber \\
 \fl \qquad \quad  + \sum_{AC} \cO_{A \beta C} (v_i^A u_j^{\beta } - v_j^A u_i^{\beta }) v_k^C 
+ \sum_{A} \cP_{A \beta \gamma } (v_i^A u_j^{\beta } - v_j^A u_i^{\beta }) u_k^{\gamma }  
\nonumber \\
\fl \qquad \quad +  \sum_{C} \cR_{\alpha  \beta C} (u_i^{\alpha } u_j^{\beta } 
- u_j^{\alpha } u_i^{\beta }) v_k^C +\cS_{\alpha  \beta \gamma } (u_i^{\alpha } u_j^{\beta } 
- u_j^{\alpha } u_i^{\beta }) u_k^{\gamma }   , \nonumber
\EEA
where $\cA, \cC, \cM, \cN, \cR$ and $\cS$ are antisymmetric in first two indices.
Now we need 
to rewrite some of the equations from this section
in terms of these new quantities (more complicated equations are given in \ref{appdecomp}).

From (\ref{eqPsi_i}) and (\ref{AS1}), it follows that
\BEA
a_{A} &=& 4 \sum_B \cM_{ABB}+ 2 \cP_{A \beta \beta } ,  \label{tra}\\
 b_{\alpha } &=& -2 \sum_B \cO_{B\alpha B} +4\cS_{\na\nb\nb} ,      \label{trb}
\EEA
and 
\BEA
(L-4\nl_A)a_{A} &=&-8 \sum_B \nl_B\cM_{ABB},  \label{tra2}\\
\ \ \ \ \ \ \ \ \ \ \ L b_{\alpha } &=& 4 \sum_B\nl_B \cO_{B\alpha B}  ,      \label{trb2}
\EEA
respectively. Equation (\ref{BS1}) implies
\BEA
L\cS_{\alpha  \beta \gamma } = 0 \label{CBS11} ,\\
(L-2\lambda_C) \cR_{\alpha  \beta C} = 0 \label{CBS12} ,\\
L \cP_{B \alpha \gamma } + 2 \lambda_B \cP_{B \gamma \alpha } = 0 \label{CBS13} ,\\
(L-2\lambda_C) \cO_{A \beta C}  + 4 \lambda_A \cN_{AC \beta } + \lambda_A b_{\beta } \delta_{AC} = 0 \label{CBS14} , \\
 2 \lambda_A \cO_{A \gamma B} - 2 \lambda_B \cO_{B\gamma A} + 2L \cN_{AB\gamma } = 0 \label{CBS15} ,\\
 (L - 2 \lambda_C) \cM_{ABC} + 2 \lambda_A \cM_{ACB} + 2\lambda_B \cM_{CBA} + \nl_C \nd_{C[A}a_{B]} =0,\label{CBS16} 
 \EEA
and (\ref{eqPsicykl}) leads to
\BEA
\cP_{A[\beta \gamma ]} +  \cR_{\beta \gamma A} = 0 ,\label{Bia01}  \\
\cO_{[A| \beta | C]}  + \cN_{CA \beta } = 0 ,\label{Bia02} \\
\cM_{\{ ABC\} }  = 0 ,\label{Bia03} \\
\cS_{\{ ABC\} }  = 0 .\label{Bia04} 
\EEA


In the twisting case, the equations (\ref{Bia33}) and (\ref{Bia35}),
without specifying dimension, are quite complex, and there are too many distinct cases to be solved. 
We thus present the solution provided $L \not=0$ and  provided that for every pair $A$, $C$ ($A\not= C$)
there exists $\beta$ for which $\cO_{A\beta C}\not= 0$. It turns out that the only eigenvalues of $\mS$
that are compatible with (\ref{Bia33}) and (\ref{Bia35}) correspond to $\Am=2$, $\nl_1=L/2$, $\nl_2=L/2$.
We also checked  other cases (but not all of them) and they also lead to the same conclusion.
For completeness, we treat all other possible cases in five dimensions in \ref{app5d}.

A linear combination of (\ref{CBS15}) and (\ref{Bia02}), [(\ref{CBS15}) - $L$ (\ref{Bia02})], gives
\BE
(2 \nl_A + L) \cO_{A \gamma B} = (2 \nl_B + L) \cO_{B \gamma A}. \label{eqO1}
\EE
For $A\not=B$ [(\ref{CBS14})-$2 \nl_A$ (\ref{Bia02})] leads to
\BE
(L + 2 \nl_A -2 \nl_B) \cO_{A \gamma B}= 2 \nl_A \cO_{B \gamma A} \mid_{A\not=B}, \label{eqO2}
\EE
and [$2\nl_A$ (\ref{eqO1})$-(2\nl_B+L)$(\ref{eqO2})] gives
\BE
(\nl_A^2 + \nl_B^2 - \nl_A \nl_B - \frac{L^2}{4}) \cO_{A\gamma B} = 0 \mid_{A\not=B} .  \label{eqlambdaO}
\EE

If  $\cO_{A \gamma B} \not=0$, we obtain from (\ref{eqlambdaO})  
\BE
(\nl_A^2 + \nl_B^2 - \nl_A \nl_B - \frac{L^2}{4}) = 0 \mid_{A\not=B} .  \label{eqlambda}
\EE
If for each pair $A$, $B$ there exists $\gamma$ such that $\cO_{A\gamma B}\not= 0$ then
(\ref{eqlambda}) is valid for all $A$, $B$, $A\not= B$.
In the case $\Am=2$, we have $L=\nl_1+\nl_2$ and  equation (\ref{eqlambdaO}) leads
to $(\nl_1-\nl_2)^2=0$ and consequently $\nl_1=\nl_2=\frac{L}{2}$.

In the case $\Am>2$, by subtracting
\BE
(\nl_A^2 + \nl_C^2 - \nl_A \nl_C - \frac{L^2}{4}) = 0 \mid_{A\not=C}   \label{eqlambdaC}
\EE
from (\ref{eqlambda}) we obtain
\BDM
(\nl_B - \nl_C) (\nl_B + \nl_C - \nl_A) = 0 \mid_{A\not=B, A\not=C,B\not=C}   
\EDM
and thus 
\BE
\nl_B - \nl_C=0  \label{equaleigenvals}
\EE
or 
\BDM
\nl_B + \nl_C - \nl_A \mid_{A\not=B, A\not=C,B\not=C}=0.
\EDM
However, in the second case one can show that simply by summing this equation with interchanged
indices
$$
(\nl_B + \nl_C - \nl_A =0) + (\nl_B + \nl_A - \nl_C =0) \Rightarrow \nl_A  =0
$$
which cannot happen as $\nl_A  \not=0$ by definition and thus (\ref{equaleigenvals}) is satisfied
and by substituting $\nl_A=\nl_B=\nl$ in (\ref{eqlambda}) we obtain
$
\lambda=\pm \frac{L}{2} .
$ 
Note however that $\sum_A \nl_A = L$ and thus the case $\nl_A=-L/2$ is excluded and
$\Am=2$.
Thus we can conclude with
\begin{lemma}
For vacuum type III spacetimes with $L\not=0$,
providing  that for every pair $A$, $C$ ($A\not= C$)
there exists $\beta$ for which $\cO_{A\beta C}\not= 0$, the expansion matrix
$\mS$ has just one non-vanishing eigenvalue  $\S=\frac{L}{2}$ which has multiplicity 2. 
\end{lemma}

Let us now study the case with two non-vanishing eigenvalues of $\mS$, $\nl_1=\nl_2=L/2$, in general, 
i.e. without any assumptions about  $\cO_{A \beta C}$.

\subsection{The case $\Am=2$, $\nl_1=\nl_2=L/2$ and $\Psi_i\not=0$.}

From (\ref{CBS11}) and  (\ref{CBS13}) it follows that
\BE
\cS_{\alpha\beta\gamma}=0,\ \ \cP_{A \beta \gamma } = - \cP_{A \gamma \beta}. \label{antsymP}
\EE
By assuming that the index $A\not=C$ in (\ref{CBS14}) we obtain
\BDM
\cN_{AC\beta}=0.
\EDM
Since $\cN$ is antisymmetric in the first two indices, $\cN_{AC\beta}=0$ for all combinations of indices.
By substituting $A=C$ in (\ref{CBS14}) we obtain
\BDM
b_{\beta}=0 
\EDM
and thus $\Psi_i$ is an eigenvector of $\mS$ corresponding to $L/2$.
Equation (\ref{CBS14}) now implies
\BDM
\cO_{A\beta C} = \cO_{C \beta A}
\EDM
and (\ref{trb}) implies
\BE
\sum_{A} \cO_{A \beta A} = 0. \label{sumO}
\EE
Substituting $A\not=C,B\not=C$ into (\ref{CBS16}) leads to
\BDM
\cM_{ABC}=0 \mid_{A\not=C,B\not=C} ,
\EDM
(\ref{tra}) leads to
\BE
a_A = 4 \sum_B \cM_{ABB}\ \   \Rightarrow\ \   a_1= 4 \cM_{122}, a_2=4 \cM_{211}  \label{sumM}
\EE
and (\ref{Bia01}) gives
\BE
\cP_{A\beta \gamma } + \cR_{\beta \gamma A} =0 . \label{PRrel}
\EE

Substituting the index $B=A$ into  (\ref{Bia3-bc}), summing over the index $A$ and  using (\ref{Bia3-h}), 
(\ref{sumM}), (\ref{sumO})
and (\ref{antsymP}) we arrive at
$\sum_A \cB_{A\beta} a_{A} = 0$.
Consequently,  the vector ${\bf \Phi}$
\BDM
\Phi_i = A_{ij} \Psi_j = \sum_B 2 \cA_{AB} a_B v_i^A ,
\EDM
which is orthogonal to $\Psi_i$,
is also an eigenvector of $\mS$. We thus have two orthogonal eigenvectors of $\mS$, $\Psi_i$ and $\Phi_i$. Let us
denote $\Psi_i \Psi_i$ by $\psi^2$ and $\Phi_i \Phi_i$ by $\phi^2$.
Now we  choose an orthogonal basis with these two vectors which corresponds to
\BDM
v^1_i, v^2_i  \longmapsto \frac{\Psi_i}{{\psi}}, \frac{\Phi_i}{{\phi}}.
\EDM
In this basis we will denote components corresponding to ${\Psi_i}$ and ${\Phi}_i$ by indices $P$ and $F$,
respectively.
$\mS$ now takes the form
\BE
S_{ij} = \S 
\left( \frac{\Psi_i \Psi_j}{\psi^2} +  \frac{\Phi_i \Phi_j}{\phi^2}      \right). \label{vyslS}
\EE
This equation implies $
S_{ij} S_{ij} =2\S^2
$.
Now using $\ell=2(\A +\S^2 )$ it follows from the previous equations and (\ref{eqell}) that
\BDM
A_{ik} \Phi_i = \A
\Psi_k
\EDM
and thus 
\BDM
\phi^2 = \Phi_i \Phi_i =\Phi_i A_{ik} \Psi_k = \A
\Psi_k \Psi_k = \A\psi^2.
\EDM
$A_{ij}$ as an antisymmetric matrix takes the general form
\BEAH
A_{ij} \: = &&  \cA_{PF} (\Psi_i \Phi_j - \Psi_j \Phi_i) + \cB_{P\alpha} (\Psi_i u_j^{\alpha } - \Psi_j u_i^{\alpha }) 
\\ &+& 
\cB_{F\alpha} (\Phi_i u_j^{\alpha } - \Phi_j u_i^{\alpha }) 
+ \cC_{\alpha \beta } (u_i^{\alpha } u_j^{\beta } - u_j^{\alpha } u_i^{\beta }) 
\EEAH
but from $A_{ij} \Psi_j = \Phi_i$ and $A_{ik} \Phi_i = \A
\Psi_k$
we get
\BDM
\cB_{P\alpha}=0, \ \ \cA_{PF} = -\frac{1}{\psi^2 } , \ \ \cB_{F\alpha }=0
\EDM
and thus
\BDM
A_{ij} = \frac{1}{\psi^2} (\Phi_i \Psi_j - \Phi_j \Psi_i) 
 + \cC_{\alpha \beta } (u_i^{\alpha } u_j^{\beta } - u_j^{\alpha } u_i^{\beta }) .
\EDM
Thanks to (\ref{antsymP}) -- (\ref{PRrel}) we can rewrite $\Psi_{ijk}$ in the form
\BEA
\fl
\Psi_{ijk}= \frac{1}{2\phi^2} (\Psi_i \Phi_j - \Psi_j \Phi_i) \Phi_k
+ \frac{\cO_{P\alpha P}}{\psi^2} (\Psi_i u_j^{\alpha}-\Psi_j u_i^{\alpha}) \Psi_k
- \frac{\cO_{P\alpha P}}{\phi^2} (\Phi_i u_j^{\alpha}-\Phi_j u_i^{\alpha}) \Phi_k \nonumber  \\
\fl
+ \frac{\cO_{P\alpha F}}{{\psi \phi}} (\Psi_i u_j^{\alpha}-\Psi_j u_i^{\alpha}) \Phi_k
+ \frac{\cO_{P\alpha F}}{{\psi \phi}} (\Phi_i u_j^{\alpha}-\Phi_j u_i^{\alpha}) \Psi_k 
+ \frac{\cP_{P\beta \gamma }}{ \psi} (\Psi_i u_j^{\beta} - \Psi_j u_i^{\beta}) u_k^{\gamma} \nonumber \\
\fl
+ \frac{\cP_{F\beta \gamma }}{ \phi} (\Phi_i u_j^{\beta} - \Phi_j u_i^{\beta}) u_k^{\gamma} 
- \frac{\cP_{P\beta \gamma }}{ \psi} (u_i^{\beta}  u_j^{\gamma}- u_j^{\beta}  u_i^{\gamma}) \Psi_k 
- \frac{\cP_{F\beta \gamma }}{ \phi} (u_i^{\beta}  u_j^{\gamma}- u_j^{\beta}  u_i^{\gamma}) \Phi_k .
\label{Psiform}
\EEA
By substituting (\ref{Psiform}) into (\ref{Bia5-d}) we obtain
$\cA_{PF}\cP_{P \gamma\delta } = 0$,
$\cA_{PF}\cP_{F \gamma\delta } = 0$
and since $\cA_{PF}\not= 0$ we also get
$\cP_{P \gamma \delta } = 0 = \cP_{F \gamma \delta }$.
Then (\ref{Bia5-f}) or (\ref{Bia5-eg}) gives  $\cC_{\alpha\beta}\cM_{CDE}=0$
which, since  $\cM_{CDE}$ does not vanish,  implies $\cC_{\alpha\beta}=0$.
In this case we thus have
\BE
A_{ij} = \frac{1}{\psi^2} (\Phi_i \Psi_j - \Phi_j \Psi_i), \label{vyslA}
 \EE
and 
 \BEA
\Psi_{ijk}&=& \frac{1}{2\phi^2} (\Psi_i \Phi_j - \Psi_j \Phi_i) \Phi_k \nonumber \\
&+& \frac{\cO_{P\alpha P}}{\psi^2} (\Psi_i u_j^{\alpha}-\Psi_j u_i^{\alpha}) \Psi_k
- \frac{\cO_{P\alpha P}}{\phi^2} (\Phi_i u_j^{\alpha}-\Phi_j u_i^{\alpha}) \Phi_k \nonumber  \\
&+& \frac{\cO_{P\alpha F}}{{\psi \phi}} (\Psi_i u_j^{\alpha}-\Psi_j u_i^{\alpha}) \Phi_k
+ \frac{\cO_{P\alpha F}}{{\psi \phi}} (\Phi_i u_j^{\alpha}-\Phi_j u_i^{\alpha}) \Psi_k .  \label{vyslB}
\EEA
We do not need to examine the rest of equations in \ref{appdecomp} since $S_{ij}, A_{ij}$ and $\Psi_{ijk}$ 
given by (\ref{vyslS}), (\ref{vyslA}) and (\ref{vyslB}), respectively,
already satisfy both equations (\ref{Bia33}) and (\ref{Bia35}) and thus represent  their solution. 

Note that from (\ref{vyslS}), (\ref{vyslA}) and (\ref{vyslB}) we obtain for type III spacetimes
relations equal or analogous to equations given in Lemma \ref{lem}:
\BEA
(a)\ \ && \Psi_{ijs}S_{sk}=\S \Psi_{ijk} ,\\
(b)\ \ && S_{ik}S_{jk}=\S S_{ij} ,\\
(c)\ \ && A_{ik}S_{kj}=S_{ik}A_{kj}=\S A_{ij} ,\\
(d)\ \ && \S A_{ik}A_{kj}= -\A 
S_{ij} ,\\
(e)\ \ && \S 
L_{ik}L_{jk}=\S L_{ki}L_{kj} = (\S^2 
+\A ) S_{ij}  .
\EEA

\section{Comments on a possible generalization of the Goldberg--Sachs theorem
for higher dimensions}
\label{Goldberg}

In Sections \ref{typeN} and \ref{PIII}, it is shown that for type N and III vacuum spacetimes the expansion
and twist matrices
$\mS$ and $\mA$ have very specific properties given by (\ref{decompPsiS}), (\ref{decompA})  and
(\ref{vyslS}), (\ref{vyslA}) for type N and III, respectively. Note, however, that while in higher 
dimensions as well as in 4D the multiple PND is geodesic, it is not shearfree for $n>4$.
The question thus arises whether there exist some properties of matrices $\mS$ and $\mA$
that are satisfied (together with the condition that the spacetime possesses a geodesic null
congruence) if and only if the vacuum spacetime is algebraically special.
The answer is unclear at present. Moreover, the conditions for $\mS$ and $\mA$ 
that hold for types N and III are not satisfied for type D spacetimes
(see \ref{typeD}). Let us here, as a first step towards such possible generalization,
show that from the Ricci identities it follows that
\begin{lemma}
\label{lemGS}

 a)
Suppose that an otherwise arbitrary vacuum spacetime admits a non-expanding and non-twisting geodesic null congruence
(i.e. $\mS=0=\mA$). Then, the spacetime in question  is algebraically special.\\
b)
Suppose that an otherwise arbitrary vacuum spacetime admits an expanding and twisting geodesic null congruence,
and that its $\mS$ and $\mA$ matrices, in appropriately chosen frame, have the form
(\ref{decompPsiS}), (\ref{decompA}). Then, the spacetime in question  is algebraically special.

\end{lemma}



\begin{proof}
The contraction of the Ricci identities $\ell_{a;bc}-\ell_{a;cb}-{R^s}_{abc}\ell_s=0$ with 
$m^{(i)a}\ell^b m^{(j)c}$ assuming that $\bl $ is geodesic with an affine parametrization
leads to
\BE
DL_{ij}+L_{is}L_{sj}+L_{is}\Ms_{j0}+L_{sj}\Ms_{i0}+R_{0i0j}=0 .\label{Riccilmm}
\EE
If $\mL=\mS+\mA=0$ then $R_{0i0j}=C_{0i0j}=0$ 
for all $i$, $j$ and the spacetime 
is thus algebraically special.
This proves  the part (a) of  Lemma \ref{lemGS}.

If we switch $i$ and $j$ in (\ref{Riccilmm})
and add and subtract the two equations we obtain
\BEA
2DS_{ij}+L_{is}L_{sj}+L_{js}L_{si}+2S_{is}\Ms_{j0}+2S_{sj}\Ms_{i0}+2R_{0i0j}=0 ,\label{RicciS}\\
2DA_{ij}+L_{is}L_{sj}-L_{js}L_{si}+2A_{is}\Ms_{j0}+2A_{sj}\Ms_{i0}=0 .\label{RicciA}
\EEA

We further assume that we have chosen vectors $\bm^{(i)}$ in such a way that
${\bf S}$ is diagonal with the only non-zero entries being $S_{22}=S_{33}$
and that $\mA$ has only two non-vanishing components $A_{23}=-A_{32}$. 
Then (\ref{RicciS}) takes the form
\BE
2DS_{ij}+\frac{2}{\S}
(\S^2-\A)S_{ij}+2S_{is}\Ms_{j0}+2S_{sj}\Ms_{i0}+2R_{0i0j}=0 .
\label{RicciSs}
\EE
If $i,j>3$ then $R_{0i0j}=0$. If $i\not=j <4$ then (\ref{RicciSs}) implies
$2S_{22}{\stackrel{2}{M}}_{30}+2S_{33}{\stackrel{3}{M}}_{20}+2R_{0203}=0$ which using 
(\ref{const-scalar-prod}) leads to
$R_{0203}$. If $i=j<4$ then again from (\ref{RicciSs}) we get 
$2D\S +2(\S^2-\A )+2R_{0202}=2D\S+2(\S^2 -\A)+2R_{0303}=0$ which
together with $R_{0i0i}=0$ gives $R_{0202}=R_{0303}=0$. For $j>3$ and $i\in (2,3)$
 equations (\ref{RicciSs})  and (\ref{RicciA}) turn to be
\BEA
\!\!\!\!\!\!\!\!\!\!\!\!\!\!\!\!\!\!\!\!\!\!\!\!\!\!\!
2S_{is}\Ms_{j0}+2R_{0i0j}=0\ \ \Longrightarrow\ \ S_{22}{\stackrel{2}{M}}_{j0}+R_{020j}=0,\  
S_{33}{\stackrel{3}{M}}_{j0}+R_{030j}=0,\  \\
\!\!\!\!\!\!\!\!\!\!\!\!\!\!\!\!\!\!\!\!\!\!\!\!\!\!\!
2A_{is}\Ms_{j0}=0\ \ \Longrightarrow  A_{23}{\stackrel{3}{M}}_{j0}=A_{32}{\stackrel{2}{M}}_{j0}=0\ 
\ \Longrightarrow \ 
{\stackrel{2}{M}}_{j0}={\stackrel{3}{M}}_{j0}=0
\EEA
that leads to $R_{020j}=R_{030j}=0$ and hence  $R_{0i0j}=C_{0i0j}=0$ for all $i$, $j$ and the spacetime 
is thus algebraically special. This proves  the part (b) of  Lemma \ref{lemGS}.
\end{proof}

Note also that by contracting  $i$ with $j$ in (\ref{RicciSs}) we obtain for higher dimensions
the same relation which is also valid for 4D
\BE
D\S =\A-\S^2 .
\EE

\section{Discussion}
\label{secdisc}

In this paper, we  present a higher dimensional frame formalism. The complete set of frame
components of the Bianchi identities, which  are in this context 
usually called the Bianchi equations, is given in \ref{appbianchi}.
 For algebraically general
spacetimes these equations are quite complicated.  However, for algebraically special cases they are much simpler 
(e.g., see Section 3 for
the type N case).  In 4D it is possible to use the Bianchi and Ricci equations to construct many algebraically special
solutions of Einstein's field equations. The hope is that it is possible to do a similar thing in higher dimensions,
at least for the simplest algebraically special spacetimes. The vast majority of today's known 
higher dimensional exact solutions are simple
generalizations of 4D solutions. The present  approach may lead to new, genuinely higher dimensional exact solutions.
Type N and D (see \cite{Algclass,Weylletter} for the definition of  type D in higher dimensions)
solutions may be of particular physical interest. 

In particular, we  study the consequences of the algebraic  Bianchi equations for type N
and III spacetimes. It turns out that the principal null direction $\bl$ in these spacetimes is geodesic.
For vacuum type N spacetimes with non-vanishing expansion or twist 
we also prove that  the corresponding components of the
Weyl tensor and expansion and twist matrices, in an appropriately chosen  frame,
can be
expressed as 
\BDM
\fl
\Psi_{ij} = \sqrt{\frac{p}{2}}  (v_i v_j - w_i w_j), \quad
S_{ij} = \S 
 (v_i v_j + w_i w_j) , \quad
A_{kl} = {A}
(w_k v_l  - v_k w_l).
\EDM
Note that we do not obtain any constraints for $\Psi_{ij}$ in non-twisting and non-expanding
type N spacetimes. 
We also establish similar results for vacuum type III spacetimes.

The Weyl tensor is said to be reducible if we can decompose it into two parts
$$
C^a_{\ bcd} = C^{\tilde a}_{\ \tilde b \tilde c \tilde d} + C^{\hat a}_{\ \hat b \hat c \hat d}
$$
where indices $a,b, \dots$ have values from 0 to $n-1$, whereas indices 
$\tilde a, \tilde b, \dots$ have values from 0 to $N-1$, and indices $\hat a, 
\hat b, \dots$  from N to $n-1$ \cite{Weylletter}.  
In this sense the Weyl tensor in vacuum type N spacetimes with non-vanishing expansion 
or twist is reducible, with a nontrivial four-dimensional part and  a vanishing $(n-4)$-dimensional part.

In 4D the well-known Goldberg--Sachs theorem \cite{GS} states that `a vacuum metric is algebraically special 
if and only if it contains a shearfree geodesic null congruence'. This theorem (and also its non-vacuum generalizations -- 
see \cite{Stephani}) is very useful for constructing algebraically special
exact solutions. At present it is unclear if and 
to what extent this theorem may be generalized for higher dimensions
(see Sec. \ref{Goldberg} for details).

Recently, all 4D spacetimes with vanishing curvature invariants (i.e., vanishing invariants
constructed from the Riemann tensor
and its covariant derivatives of an arbitrary order -- VSI spacetimes) were determined in  
\cite{VSI}. In \cite{PP6}, the generalization to higher dimensions was discussed. All these
spacetimes are of type III, N or O.
The results presented in this paper will enable us to explicitly express   curvature
invariants involving derivatives of the Riemann tensor in higher dimensions.
For example, for type N vacuum spacetimes we  obtain, thanks to Lemma 1,
\BDM
I=C^{a b c d  ; pq} C_{a m c  n ; pq} 
C^{t m u n ; rs} C_{t b u d  ; rs} = 3^2 2^{10} p^2 (S^2+A^2)^4.
\EDM
Consequently, we will be able 
to prove that higher dimensional VSI spacetimes are expansion and twist free, thereby
proving the assertion made in \cite{PP6} (see \cite{VSID}). 

\ack
We are grateful to the referee for constructive criticism.
AP and VP would like to thank Dalhousie University for its hospitality while part of this work was carried out.
RM and AC were partially supported by a research grant from NSERC. VP was supported by grant GACR-202/03/P017 and
AP by grant KJB1019403.
AP and VP also thank to N\v CLF for support.

\appendix
\section{Components of equations (\ref{Bia33}) and (\ref{Bia35})}
\label{appdecomp}

Equation (\ref{Bia33}) gives
\BEA
\fl
\cC_{\na[\nb}b_{\ng ]}+\sum_F \cB_{F[\nb |}\cP_{F\na |\ng]}+4\cC_{\phi [\nb |}\cS_{\phi\na |\ng]}=0,
\label{Bia3-h} 
\\ 
\fl
2 \cC_{\na \nb} a_C + \cB_{C\na} b_{\nb} -2 \nl_C \cP_{C\na\nb} + \sum_F 4\cA_{CF} \cP_{F\na\nb} 
+ 2 \sum_F \cB_{F\nb} \cO_{F\na C} \nonumber \\ 
+ 8 \cC_{\nf \nb} \cR_{\nf \na C} + 4 \cB_{C \nf} \cS_{\nf \na \nb}=0,
\label{Bia3-fg} \\ 
\fl
 \cB_{A [\nb} b_{\ng]} + 4 \sum_F \cB_{F [\nb|} \cN_{FA|\ng]} - 4 \cC_{ \nf [\nb|} \cP_{ A\nf  |\ng]} = 0,
\label{Bia3-e} \\ 
\fl
 \nl_B \cO_{B \na C} -  \nl_C \cO_{C \na B} + \cB_{[C |\na} a_{|B]} + 4 \sum_F \cA_{F[B|} \cO_{F \na |C]} 
+ 4 \cB_{[B| \nf} \cR_{\na \nf |C]}=0 ,\label{Bia3-d}  \\ 
\fl
\nl_A \nd_{AB} b_{\ng} - \cB_{A \ng} a_B + 2 \cA_{AB} b_{\ng} + 4 \nl_B \cN_{BA \ng} + 8 \sum_F \cA_{FB} \cN_{FA \ng}
 \nonumber \\ 
\fl
+2 \cB_{B \nf} \cP_{A \nf \ng} - 4 \sum_F \cB_{F \ng} \cM_{FAB} + 4 \cC_{\nf \ng} \cO_{A \nf B} = 0,
\label{Bia3-bc} \\ 
\fl
 \nl_{A} \nd_{A[B} a_{C]} + 2 \cA_{A[B}a_{C]} + 2 \nl_B \cM_{BAC} - 2 \nl_C \cM_{CAB} 
\nonumber \\ + 8\sum_F  \cA_{F [B|} \cM_{FA|C]}+2\cB_{[ B| \nf}\cO_{A\nf |C]}=0.\label{Bia3-a} 
\EEA

Equation (\ref{Bia35}) leads to
\BEA
\fl
 2\cC_{\{ \na\nb|} \cS_{\ng\nd|\ne\} }+\cC_{\nd\{ \na} \cS_{\nb\ne\} \ng}-\cC_{\ng\{ \na} \cS_{\nb\ne\} \nd}=0,
\label{Bia5-n} \\
\fl
 2 \cC_{\nb \ne} \cR_{\ng \nd A} +  \cC_{\nd [\ne|} \cP_{A |\nb] \ng}
+  \cC_{\ng [\nb|} \cP_{A |\ne] \nd} + 2 \cB_{A [\nb} \cS_{\ng \nd |\ne]} 
+ \cB_{A [\ng} \cS_{\nb \ne |\nd]} = 0, 
\label{Bia5-m} \\
\fl 
 2 \cC_{ \{\na \nb|} \cP_{C \nd| \ne\}} + 2 \cC_{\nd \{\na} \cR_{\nb \ne\} C} 
- \cB_{C \{\na} \cS_{\nb \ne\} \nd}  = 0, 
\label{Bia5-l} \\
\fl
-  4 \cB_{[A| \ne} \cR_{\ng \nd |B]} +  \cB_{[B|\nd} \cP_{|A] \ne \ng} - \cB_{[B|\ng} \cP_{|A] \ne \nd}
+ 4 \cC_{[\nd| \ne} \cN_{AB|\ng]} +4 \cA_{AB}\cS_{\ng\nd\ne}= 0 ,\label{Bia5-k} \\
\fl
2 \cB_{A [\nb|} \cP_{C \nd |\ne]} +  \cB_{C [\nb|} \cP_{A  | \ne] \nd} + 2\cC_{\nb \ne} \cO_{C \nd A}
- 2 \cC_{\nd [\nb|} \cO_{A |\ne] C} - \cB_{A \nd} \cR_{\nb \ne C} \nonumber \\ 
-(\nl_A\nd_{AC}+2\cA_{CA}) \cS_{\nb\ne\nd}= 0,\label{Bia5-ij}  \\
\fl
4 \cC_{\{\na \nb|} \cN_{CD|\ne\}} +  \cB_{D \{\na} \cR_{\nb \ne\} C} -  \cB_{C \{\na} \cR_{\nb \ne\} D} = 0 ,
\label{Bia5-h} \\
\fl
4 \cC_{\na \nb} \cM_{CDE} + 4 \cB_{E [\na|} \cN_{CD|\nb]} + \cB_{D[\nb|} \cO_{E |\na] C}
-  \cB_{C[\nb|} \cO_{E |\na] D}  \nonumber \\ 
+ 2 \nl_E  \nd_{E[D} \cR_{\na \nb |C]}+ 4 \cA_{[D|E} \cR_{\na \nb |C]} = 0, \label{Bia5-eg} \\
\fl
2 \cC_{\na \ng} \cM_{BED} +  \cB_{D \na} \cN_{BE\ng} + 2 \cB_{[B|\na} \cO_{D \ng |E]} 
+  \cB_{[B|\ng} \cO_{E \na |D]} \nonumber \\
\fl
- 2 \cA_{BE} \cP_{D\ng \na} + 2 \cA_{D[B} \cP_{E]\na\ng} + \nl_D \nd_{D[B} \cP_{E]\na \ng} = 0, 
\label{Bia5-f} \\
\fl
 \cB_{[A| \ng} \cN_{B]E\nd} +  \cB_{[B| \nd} \cN_{A]E\ng} +\cB_{E[ \ng }\cN_{AB| \nd ]}
+ 2 \cA_{\{AB|} \cR_{\ng \nd |E\}} 
= 0, \label{Bia5-d} \\
\fl
4 \cB_{[E|\na} \cM_{CD|B]} + 2 \cB_{[D|\na} \cM_{BE|C]} + 4 \cA_{BE} \cN_{CD \na} +2 \cA_{D[B} \cO_{E]\na C}
+ 2 \cA_{C[E} \cO_{B]\na D} \nonumber \\
\fl
+ \nl_B \nd_{B[D|} \cO_{E\na |C]}  + \nl_E \nd_{E[C|} \cO_{B\na |D]}  = 0,\label{Bia5-c} \\
\fl
\cB_{\{ A |\ng}\cM_{BE\} D}-2\cA_{\{ AB|}\cO_{D\ng |E\}}+2\cA_{D\{ A}\cN_{BE\}\ng}+\nl_D\nd_{D\{ A} \cN_{BE\} \ng}=0,
\label{Bia5-b} \\
\fl
4\cA_{\{ AB|} \cM_{CD|E\} }+2\cA_{D\{ A} \cM_{BE\} C}-2\cA_{C\{ A} \cM_{BE\} D} \nonumber \\
 +\nl_D\nd_{D\{ A} \cM_{BE\} C}
-\nl_C\nd_{C\{ A} \cM_{BE\} D}=0.\label{Bia5-a} 
\EEA

\section{Bianchi equations}
\label{appbianchi}

\BEA
\fl
 D R_{101i} + \T R_{010i}-\delta_i R_{0101}= 
2R_{010[i} L_{1]1}- R_{101i} L_{10}
\nonumber\\
\fl
- R_{01is} L_{s1}- 2R_{101s} L_{si} 
+  2R_{[1|i1s} L_{s|0]}    
- R_{0101} N_{i0}
  - 2R_{010s} N_{si} 
\nonumber\\
\fl
 +  R_{01is} N_{s0} +  2R_{0s[0|i} N_{s|1]}    
- R_{010s} \Ms_{i1} - R_{101s} \Ms_{i0}  ,
\EEA

\BEA
\fl
- \T R_{01ij}+2\delta_{[j|}R_{101|i]} =
2R_{101[j|} L_{1|i]}  +  2R_{101[i|} L_{|j]1} +  2R_{1[i|1s} L_{s|j]}+  R_{1sij} L_{s1}\nonumber\\
\fl
 +  2R_{0101} N_{[ji]}+  2R_{010[j} N_{i]1}  +  2R_{01[i|s} N_{s|j]} +  2R_{0s1[j|} N_{s|i]} - R_{0sij} N_{s1}  
\nonumber \\
\fl
  +  2R_{01[i|s} \Ms_{|j]1}    
 +  2R_{101s} \Ms_{[ji]}   ,
\EEA
\BEA
\fl
-D R_{01ij} +2\delta_{[i|}R_{010|j]}=  2R_{0101} L_{[ij]}+2R_{010[j|} L_{1|i]} +  2R_{01[i|s} L_{s|j]}     
\nn\\\fl
+2R_{0[i|1s} L_{s|j]}  +  2R_{101[i} L_{j]0} +  R_{1sij} L_{s0}  +  2R_{010[j} N_{i]0} +  2R_{0[j|0s} N_{s|i]}   
 \nn \\
\fl
 - R_{0sij} N_{s0}  +  2R_{01[i|s} \Ms_{|j]0}  +  2R_{010s} \Ms_{[ij]}  ,
\EEA
\BEA
\fl
D R_{1i1j}-\T R_{0j1i}-\delta_j R_{101i}=
2R_{101i} L_{[1j]}- 2R_{1i1j} L_{10}+  2R_{1i[j|s} L_{s|1]}  \nn\\
\fl  
+  2R_{010[1|} N_{i|j]}- 2R_{101(i|} N_{|j)0} +  2R_{0[j|is} N_{s|1]}+  R_{0s1i} N_{sj}   
- 2R_{1(ij)s} N_{s0}  \nn\\
\fl
+ 2 R_{[0|j1s} \Ms_{i|1]}+ 2 R_{[0|s1i} \Ms_{j|1]}  +  R_{101s} \Ms_{ij},      
\EEA
\BEA
\fl
D R_{0i1j}-\T R_{0i0j}+\delta_j R_{010i}=   
2R_{010[j|} L_{i|1]}+  2R_{010i} L_{(1j)}  - 2R_{0i0j} L_{11}\nn\\ 
\fl
+  R_{101j} L_{i0}+  2R_{0i[j|s} L_{s|1]}+2R_{0[j|is} L_{s|1]}- R_{1jis} L_{s0}   
+  R_{010i} N_{j0}+  2R_{0i[0|s} N_{s|j]}\nn\\
\fl
- R_{010s} \Ms_{ij} +2 R_{0s[0|j} \Ms_{i|1]}+2 R_{0i[0|s} \Ms_{j|1]}  ,
\EEA

\BEA
\fl
-D R_{1kij}+\T R_{0kij}-\delta_k R_{01ij}=
  - 2R_{0[1|ij} L_{|k]1}  +  2R_{0k1[i} L_{j]1}  +  2R_{101[i} L_{j]k} \nn\\
\fl
- 2R_{1[i|1k} L_{|j]0} +  R_{1kij} L_{10}  + 2 R_{[1|sij} L_{s|k]}
- 2R_{010[i} N_{j]k}- R_{01ij} N_{k0} 
\nn\\
\fl
+  2R_{0[i|0k} N_{|j]1}  +  2R_{0[j|1k} N_{|i]0}  +  2R_{[k|sij} N_{s|0]}
\nn \\ \fl 
 +  2R_{01[i|s} \Ms_{|j]k}  +  2R_{[0|kjs} \Ms_{i|1]}  +  
2R_{[1|kis} \Ms_{j|0]}  +  2R_{[1|sij} \Ms_{k|0]},  
\EEA
\BEA
\fl
\T R_{0ijk}+2\delta_{[k|} R_{0i1|j]}= 2R_{0i1[j|} L_{|k]1}+  2R_{0[i|jk} L_{|1]1}  +  2R_{101[j|} L_{i|k]}    
+  2R_{1[k|is} L_{s|j]}  \nn\\
\fl
- R_{isjk} L_{s1} +  2R_{010i} N_{[jk]}
+2R_{0i0[j} N_{k]1} +  2R_{0i[k|s} N_{s|j]} 
\nn \\ \fl
+  2R_{0i1s} \Ms_{[kj]} +  2R_{0s1[k|} \Ms_{i|j]}  
  +  2R_{0i[k|s} \Ms_{|j]1} - R_{0sjk} \Ms_{i1} ,
\EEA
\BEA
\fl
D R_{0ijk}+2\delta_{[k|}R_{0i0|j]}=- 2R_{010i} L_{[jk]}+2R_{010[k|} L_{i|j]}    
 +  4R_{0i0[j|} L_{1|k]}+  2R_{0i1[j} L_{k]0}\nn\\
\fl  
+  2R_{0[i|jk} L_{|1]0}+  2R_{0i[k|s} L_{s|j]}  +  2R_{0[k|is} L_{s|j]} - R_{isjk} L_{s0}   
+  2R_{0i0[j|} N_{|k]0} 
\nn \\ \fl
 +  2R_{0i0s} \Ms_{[kj]}   
+ 2 R_{0[k|0s} \Ms_{i|j]}  +2  R_{0i[k|s} \Ms_{|j]0} - R_{0sjk} \Ms_{i0}  ,
\EEA
\BEA
\fl
DR_{1ijk}+2\delta_{[k|}R_{0|j]1i}=2R_{101i} L_{[jk]}      
+  2R_{1i1[j} L_{k]0}- R_{1ijk} L_{10}+  2R_{1i[k|s} L_{s|j]}  \nn\\
\fl
+  R_{010j} N_{ik} +  2R_{01[j|k} N_{i|0]} - 2R_{0[k|1i} N_{|j]0}+  2R_{0[k|is} N_{s|j]}- R_{isjk} N_{s0}      
\nn\\ \fl
+  2R_{0[k|1s} \Ms_{i|j]}   - 2R_{0s1i} \Ms_{[jk]}+  2R_{1i[k|s} \Ms_{|j]0}- R_{1sjk} \Ms_{i0},
\EEA
\BEA
\fl
\T R_{1ijk}+2\delta_{[k|}R_{1i1|j]}=
  2R_{1i1[j} L_{k]1}+  4R_{1i1[k|} L_{1|j]}- R_{1ijk} L_{11}  \nn\\
\fl
 +  R_{01jk} N_{i1}  +2R_{0[j|1i} N_{|k]1}  +  2R_{101[k|} N_{i|j]}  +  2R_{101i} N_{[kj]}  
+  2R_{1i[k|s} N_{s|j]}+  2R_{1[k|is} N_{s|j]}  \nn \\ \fl
- R_{isjk} N_{s1} 
- 2R_{1i1s} \Ms_{[jk]}  +  2R_{1[k|1s} \Ms_{i|j]}  +  2R_{1i[k|s} \Ms_{|j]1}  
- R_{1sjk} \Ms_{i1} ,
\EEA
\BEA
\fl
-\delta_{\{i|}R_{01|jk\}}=
R_{101\{i} L_{jk\}}  - R_{101\{i} L_{kj\}}  
+  R_{1s\{ij|} L_{s|k\}} 
+R_{010\{i} N_{kj\}}    \nn\\ 
\fl
- R_{010\{i} N_{jk\}}- R_{0sij} N_{sk}+  2R_{0s[i|k} N_{s|j]} 
+  R_{01\{i|s} \Ms_{|jk\}} - R_{01\{i|s} \Ms_{|kj\}}  ,
\EEA

\BEA
\fl
-DR_{ijkm}+2\delta_{[k|}R_{0|m]ij}=
2R_{01ij} L_{[km]}  +  2R_{0k[1|j} L_{|i]m} +  2R_{0[m|1i} L_{j|k]} \nn\\
\fl
+  2R_{0m[i|j} L_{|1]k} 
+  2R_{1[j|km} L_{|i]0}+  2R_{1[m|ij} L_{|k]0}+  2R_{ij[k|s} L_{s|m]} 
\nn\\
\fl
  +  2R_{0i0[m|} N_{j|k]} 
+  2R_{0j0[k|} N_{i|m]} +  2R_{0[j|km} N_{|i]0}  
+  2R_{0[m|ij} N_{|k]0}    +  2R_{0[k|is} \Ms_{j|m]}
\nn\\
\fl
+  2R_{0[m|js} \Ms_{i|k]}  
  +  2R_{0sij} \Ms_{[km]}  +  2R_{[i|skm} \Ms_{|j]0}  +  2R_{ij[k|s} \Ms_{|m]0} , 
\EEA
\BEA
\fl
-\T R_{ijkm}+2\delta_{[k|}R_{1|m]ij}=
2R_{1i[1|m} L_{j|k]} +  2R_{1[j|1k} L_{|i]m}  
+2R_{1j[k|m} L_{i|1]} \nn \\ \fl
+  2R_{1kij} L_{[1m]}   +  2R_{1mij} L_{[k1]}  
- 2R_{01ij} N_{[km]}   +  2R_{0[i|1m} N_{|j]k}
 +  2R_{0[j|1k} N_{|i]m}\nn \\ \fl
+  2R_{0[j|km} N_{|i]1}  +  2R_{0[m|ij} N_{|k]1}
  +  2R_{ij[k|s} N_{s|m]}  
+  2R_{1k[i|s} \Ms_{|j]m} \nn \\ \fl
 +  2R_{1m[j|s} \Ms_{|i]k}  
+  2R_{1sij} \Ms_{[km]} +  2R_{ij[k|s} \Ms_{|m]1} +  2R_{[i|skm} \Ms_{|j]1}    ,
\EEA
\BEA
\fl
\delta_{\{j|}R_{1i|mk\}}=  
R_{1i1\{j} L_{mk\}} - R_{1i1\{j} L_{km\}} +  R_{1i\{jk|} L_{1|m\}} 
+  R_{01\{jm|} N_{i|k\}}   
- R_{0\{j|1i} N_{|km\}} \nn\\ \fl
 +  R_{0\{j|1i} N_{|mk\}} +  R_{is\{jk|} N_{s|m\}}  
- R_{1i\{j|s} \Ms_{|mk\}}
 +  R_{1i\{j|s} \Ms_{|km\}}  
+  R_{1s\{jk|} \Ms_{i|m\}} ,
\EEA
\BEA
\fl
\delta_{\{ j|}R_{0i|mk\}}=
R_{01\{jk|} L_{i|m\}}
- R_{0i1\{j} L_{km\}}  +  R_{0i1\{j} L_{mk\}} 
+  R_{0i\{jm|} L_{1|k\}}+  R_{is\{jk|} L_{s|m\}}
\nn \\ \fl         
+R_{0i0\{j} N_{mk\}} - R_{0i0\{j} N_{km\}}  
+  R_{0i\{j|s} \Ms_{|km\}}
- R_{0i\{j|s} \Ms_{|mk\}}  
+  R_{0s\{jk|} \Ms_{i|m\}},
\EEA

\BEA
\fl
\delta_{\{ k|}R_{ij|nm\}}=
R_{1j\{km|} L_{i|n\}}- R_{1i\{km|} L_{j|n\}} - R_{1\{k|ij} L_{|mn\}} +  R_{1\{k|ij} L_{|nm\}}  
\nn \\ \fl 
  +  R_{0j\{km|} N_{i|n\}} - R_{0i\{km|} N_{j|n\}} +  R_{0\{k|ij} N_{|nm\}} - R_{0\{k|ij} N_{|mn\}}
\nn \\ \fl  
 +  R_{ij\{k|s} \Ms_{|mn\}} - R_{ij\{k|s} \Ms_{|nm\}}+  R_{is\{km|} \Ms_{j|n\}}- R_{js\{km|} \Ms_{i|n\}}  . 
\EEA

\section{Proof for vacuum type III spacetimes}

\subsection{Non-twisting case in an arbitrary dimension}
\label{appIIInon}

In this case, $\mA$=0 and thus $\mL=\mS$ is symmetric (we assume $\mS\not= 0$ since in the
non-twisting and non-expanding case both sides of equations (\ref{Bia33}) and (\ref{Bia35}) vanish).

From (\ref{eqLPsi1}) we obtain
\BE
S_{ij} \Psi_j = \frac{L}{2} \Psi_i. \label{eqnt1}
\EE
Note that providing $\Psi_i \not= 0$  from (\ref{eqnt1}) 
and from (\ref{eqell}) we obtain  $L^2=2 \ell$ and thus the case $L=0$ leads to $\mS=0$.

Equation (\ref{eqPsiA}) leads to
\BE
L \Psi_{ijk} = 2 \Psi_{ijs} S_{sk}, 
\EE
which using the decomposition (\ref{ALB}) implies 
\BEA
\fl
\ \ \ \ \ \ \ \ \ \  \ \ \  L\cN_{AB\gamma}=0,\quad 
\ \ \ \ \ \ \ \ \ \ \ \  L\cP_{A\beta \gamma}=0,\quad 
\ \ \ \ \ \ \ \ \ \ \ \ L\cS_{\na\nb\ng}=0, \label{AoNPSL}\\
\fl
(L-2\nl_C) \cM_{ABC}=0,\quad (L-2\nl_C) \cO_{A\beta C}=0,\quad (L-2\nl_C) \cR_{\alpha \beta C}=0.\nn
\EEA

First study the case with $L=0$. From equations (\ref{AoNPSL}) we get
$\cM_{ABC}=0$, $\cO_{A\beta C}=0$, $\cR_{\alpha\beta C}=0$
and thus from (\ref{CBS13}) and (\ref{Bia02})
$\cP_{A\beta\gamma}=0$, $\cN_{AB\gamma}=0$
and further from (\ref{Bia5-ij})
$\cS_{\alpha\beta\gamma}=0$ which imply $\Psi_{ijk}=0$, i.e., a type N or O spacetime
and thus there are no non-trivial solutions of (\ref{Bia33}) and (\ref{Bia35}) of type III with $L=0$.

Now we may proceed to the case $L\not= 0$. Then  (\ref{AoNPSL})  and (\ref{Bia01}) imply
\BE
\cN_{AB\gamma}=0,\ \cP_{A\beta C}=0,\ \cS_{\na\nb\ng}=0 , \ \cR_{\na\nb C}=0.\label{AoNPS}
\EE
From (\ref{eqnt1}) we get
\BE
\nl_A a_A=\frac{L}{2}a_A,\ \ \ \mbox{and} \ \ b_\na=0 ,\ \label{nontwist_b} 
\EE
which leads to
\BE
a_A=0 \ \ \mbox{or}\ \ \nl_A=\frac{L}{2}.\label{aAl}
\EE

\begin{enumerate}
\item
If all $a_A\not= 0$ then all $\nl_A=L/2$  (\ref{aAl}) and thus $\Am=2$, $\nl_1=\nl_2=L/2$. 
The corresponding solution is given by (\ref{res_nontwist}).

\item 
If at least one $a_A\not=0$ then corresponding $\nl_A=L/2$ and from (\ref{CBS16}) with $B=C\not= A$ and (\ref{Bia3-a})
with $A=C\not= B$ and after interchanging $A$ with $B$ we get 
\BEA
a_A&=&\frac{2L-4\nl_B+4\nl_A}{\nl_B}\cM_{ABB}\mid_{B\not=A} , \\
a_A&=&4\frac{\nl_A}{\nl_B}\cM_{ABB}\mid_{B\not=A} .\label{aAM}
\EEA
Comparing the right hand sides we obtain $2L-4\nl_B+4\nl_A=4\nl_A$ and thus
$\nl_B=L/2$ for all $B\not = A$. However, the remaining $\nl_A=L/2$ and 
in this case again $\Am=2$, $\nl_1=\nl_2=L/2$.

\item
If all $a_A=0$ then (\ref{aAM}) implies  
\BE
\cM_{ABB}=0\label{nontwist_Mabb}
\EE 
for all $A$, $B$.
We treat separately the case with all $\cM_{ABC}=0$ and  the case with some $\cM_{ABC}\not=0$.

(A)

If all $\cM_{ABC}=0$ than there exists at least one non-vanishing
$\cO_{A\beta C}$ in order to have a type III Weyl tensor.   

For $\Am \geq 3$, from equation (\ref{Bia5-c}) for $B=C$, $C\not= E$, $C\not= D$, 
$D\not= E$  we get $\nl_C\cO_{E\na D}=0$
which implies that all $\cO_{E\na D}=0\mid_{E\not=D}$. 
Equation (\ref{CBS14}) with $C= A$, i.e., $(L-2\nl_A)\cO_{A\nb A}=0$,
implies
\BE
\cO_{A\nb A}=0\ \ \mbox{or}\ \ \nl_A=\frac{L}{2} .\label{nontwist_Oaa}
\EE
From equation (\ref{Bia5-c}) (for $B=C$, $D=E$, $C\not= E$)
\BE
\nl_C\cO_{D\na D}=-\nl_D\cO_{C\na C}\mid_{D\not=C}.\label{nontwist_OOaa}
\EE
If there exists $\nl_C\not= L/2$ then from (\ref{nontwist_Oaa}) $\cO_{C\na C}=0$
and thus from (\ref{nontwist_OOaa}) all $\cO_{D\na D}=0$ and $\Psi_{ijk}=0$, the spacetime is of the type N or O.
The other possibility that all $\nl_A=L/2$ is not compatible with $\Am\geq 3$ and so $\Am\leq 2$.

For $\Am=2$, (\ref{trb}) with  (\ref{nontwist_OOaa}) imply
\BE
\nl_2=\nl_1=\frac{L}{2}\ \ \mbox{and} \ \ \cO_{2\nb 2}=-\cO_{1\nb 1}\label{nontwist_O1122}
\EE
or $\cO_{2\nb 2}=\cO_{1\nb 1}=0$. In the second case, equation (\ref{Bia02}) implies
\BE
\cO_{2\nb 1}=\cO_{1\nb 2}\label{nontwist_O12}
\EE
and from (\ref{CBS15}) (or (\ref{Bia3-d})) we get
$\nl_2=\nl_1=L/2$.  

For $\Am=1$, from (\ref{nontwist_Oaa}) we immediately get $\cO_{1\nb 1}=0$, type N or O spacetimes.

To summarize,  the case $\cM_{ABC}=0$  ($\forall$ $A$, $B$, $C$)
always leads to $\Am=2$, $\nl_1=\nl_2=L/2$ or to type N or O spacetimes.

(B)

Let us now assume, for $\Am \geq 4$, that there exists $\cM_{ABC}\not= 0\mid_{B\not=C}$. 
Then (\ref{Bia5-a}) for $C\not= D$,
$A\not= D$, $A\not= C$, $B=D$, $C\not= E$  and $D\not= E$ 
implies $\nl_D\cM_{AEC}=0$  and thus all $\cM_{ABC}=0\mid_{B\not=C}$  which is in contradiction with
our assumptions and consequently $\Am \leq 3$.

Equation (\ref{CBS16}) with $C\not= A$ and $B\not= C$ and (\ref{Bia3-a}) gives
\BEA
\fl\ \ (L-2\nl_C+2\nl_A) \cM_{CAB} &=&(L-2\nl_C+2\nl_B) \cM_{CBA}\mid_{C\not=A,\ C\not=B}\label{McabMcba}
, \\
\nl_A\cM_{CAB}&=&\nl_B\cM_{CBA}.
\EEA
These two equations leads to
\BE
(L-2\nl_C)\cM_{CAB}=(L-2\nl_C)\cM_{CBA} 
\EE
and so either
\BE
\nl_C=\frac{L}{2}\ \ \wedge \ \ \nl_A\cM_{CAB}=\nl_B\cM_{CBA}\label{nontwist_lL2_Mabc}
\EE
or
\BE
\fl \nl_C\not= \frac{L}{2}\ \ \wedge \ \ \cM_{CAB}=\cM_{CBA}\ \ \wedge \ \ \nl_A\cM_{CAB}=\nl_B\cM_{CBA}
\label{nontwist_lnL2_Mabc}
\EE
i.e., 
\BE
\fl \cM_{CAB}=\cM_{CBA}=0\ \ \mbox{or}\ \ \cM_{CAB}=\cM_{CBA}\not= 0\ \ \wedge \ \ \nl_A=\nl_B.
\EE

Let us assume that $\Am=3$.
Let $\nl_1\not= L/2$. Then the case (a) $\cM_{123}=\cM_{132}=0$  (\ref{nontwist_lnL2_Mabc}) 
implies $\cM_{231}=0$ (\ref{Bia03}) which 
is the case $\cM_{ABC}=0$  studied above (A).
The condition (b) $\cM_{123}=\cM_{132}\not= 0$ (\ref{nontwist_lnL2_Mabc})  
also implies $\cM_{231}=0$  (\ref{Bia03}) and for $\nl_2\not= L/2$
we get $\cM_{213}=\cM_{231}=0$ (\ref{nontwist_lnL2_Mabc})  and thus also $\cM_{123}=0$ and this is again 
the case $\cM_{ABC}=0$.
However, if $\nl_2=L/2$ then $\nl_1\cM_{213}=\nl_3\cM_{231}=0$ (\ref{nontwist_lL2_Mabc}) 
and since $\cM_{231}=0$ we again obtain the case $\cM_{ABC}=0$.

If $\nl_1=L/2$ then  $\nl_2=L/2$ immediately implies $\nl_3=0$ and $\nl_2\not= L/2$ 
corresponds to the case analyzed in the previous paragraph.

$\Am<3$ again leads to  all $\cM_{ABC}=0$ (\ref{nontwist_Mabb}).

\end{enumerate}

To summarize,  the general solution of (\ref{Bia33}) and (\ref{Bia35}) in the non-twisting case
may be written, substituting (\ref{AoNPS}), (\ref{nontwist_b}), (\ref{aAM}), (\ref{nontwist_O1122})
and (\ref{nontwist_O12})     
into (\ref{ALB}), in the form 
\BEA
\Psi_{i}&=&4\cM_{122}v^1_i +4\cM_{211}v^2_i,\nn\\
S_{ij}&=&\frac{L}{2}(v^1_i v^1_j +v^2_i v^2_j),\label{res_nontwist}\\
\Psi_{ijk}&=& 2\cM_{122} (v^1_i v^2_j - v^1_j v^2_i) v^2_k +2\cM_{211} (v^2_i v^1_j - v^2_j v^1_i) v^1_k \nonumber \\
&+& \cO_{1\alpha 1}(v^1_i u_j^{\alpha}-v^1_j u_i^{\alpha}) v^1_k
- \cO_{1\alpha 1} (v^2_i u_j^{\alpha}-v^2_j u_i^{\alpha}) v^2_k \nonumber  \\
&+& \cO_{1\alpha 2} (v^1_i u_j^{\alpha}-v^1_j u_i^{\alpha}) v^2_k
+ \cO_{1\alpha 2} (v^2_i u_j^{\alpha}-v^2_j u_i^{\alpha}) v^1_k .\nn 
\EEA

\subsection{Twisting type III spacetimes in five dimensions}
\label{app5d}

In  Section  \ref{PIII},   
we studied possible non-zero eigenvalues $\nl_A$, $A=1, \dots, A_{{\rm max}}$,  of $\mS$ 
which are compatible with equations (\ref{Bia33}) and (\ref{Bia35}),
providing  that for every pair $A$, $C$ ($A\not= C$)
there exists $\beta$ for which $\cO_{A\beta C}\not= 0$. It turned
out that  then it follows from (\ref{Bia33}) and (\ref{Bia35})
that $\Am=2$ and $\nl_1=\nl_2=L/2$. Here we study all other possible cases
in 5D and show that the only  case with non-vanishing eigenvalues of $\mS$ is $\Am=2$ and $\nl_1=\nl_2=L/2$. Corresponding
form of $\Psi_{ijk}$, $\Psi_i$, $\mS$ and $\mA$ is given in Section   \ref{PIII}.

In 5D, $\Am$ can have values 0, 1, 2 or 3. We will treat these cases separately 
(the cases $\Am=0$ and $\Am=1$ can be easily solved in an arbitrary dimension).
\begin{enumerate}

\item
The case $\Am=0$ (globally) corresponds to $\mS=0$ and one of the Ricci equations gives $D {\bf A}={\bf A}{\bf A}$,
which implies (take trace of both sides) that ${\bf A}=0$. Thus in an arbitrary dimension,
as in 4D, there are no non-expanding, twisting spacetimes.

\item
If $\Am=1$, i.e., $\nl_1=L\not= 0$, then $\cM_{ABC}=0$,
$\cN_{AB\gamma}=0$ and $\cA_{AB}=0$. From (\ref{CBS11}) and
(\ref{CBS12}), it follows $\cS_{\alpha\beta\gamma}=0$ and $\cR_{\alpha\beta C}=0$. 
Equations (\ref{Bia01}) and (\ref{CBS13})
give $\cP_{A\beta\gamma}=0$. From (\ref{CBS14}) and (\ref{trb}), it follows $b_\beta=0$ and  $\cO_{A\beta A}=0$.
Thus, the case $\Am=1$ leads to a type N or O spacetime.

\item
In the case $\Am=2$, $\alm=1$ and thus, thanks to antisymmetry, 
$\cR_{\alpha\beta C}=0$ and $\cS_{\alpha\beta\gamma}=0$.
Then from (\ref{Bia01}) $\cP_{A\beta\gamma}=\cP_{A\gamma\beta}$ and  from (\ref{CBS13})
$\cP_{B\alpha\gamma}=0$ or $\nl_B=-L/2$. 

Let us proceed further assuming that at least one non-vanishing $\cO$ exists:

From (\ref{CBS14}) and (\ref{Bia02}), it follows that  $\cO_{112}=0$ is equivalent to $\cO_{211}=0$.
Similarly, from (\ref{CBS14}) and (\ref{trb}), $\cO_{111}=0$ is equivalent to $\cO_{212}=0$.

The case $\cO_{112},\ \cO_{211}\not= 0$ was analyzed in Section \ref{PIII} and leads to
$\nl_1=\nl_2=L/2$.

If $\cO_{112}=0=\cO_{211}$  and $\cO_{111},\ \cO_{212}\not= 0$ then (\ref{Bia02}) gives $\cN_{AB\gamma}=0$
and from (\ref{CBS14}) and (\ref{trb}),
\BE
b_\beta= \frac{\nl_1-\nl_2}{\nl_1 }\cO_{1\beta 1}=\frac{\nl_2-\nl_1}{\nl_2 }\cO_{2\beta 2}
=-2(\cO_{1\beta 1}+\cO_{2\beta 2}),
\EE
follows $\nl_1=\nl_2=L/2$. Next we have to analyze possible cases with $\cO_{A\beta C}=0$ $\forall \ A,\ \beta,\  C$.

Let us now assume that all $\cO$'s are zero and at least one $\cM$ is non-vanishing.
From (\ref{CBS16}) for $B=C$ and $B\not=A$ and from (\ref{tra2}) we get
\BEA
\fl\ \ \ &&\nl_2 a_1=2(3\nl_1-\nl_2)\cM_{122},\ \ (-3\nl_1+\nl_2) a_1=-8\nl_2\cM_{122},\label{a1} \\
\fl\ \ \ &&\nl_1 a_2=2(3\nl_2-\nl_1)\cM_{211},\ \ (-3\nl_2+\nl_1) a_2=-8\nl_1\cM_{211}.\label{a2}
\EEA 
If $a_1\not= 0$ (and thus also $\cM_{122}\not=0$) then the ratio of equations (\ref{a1}) gives 
$(3\nl_1+\nl_2)(\nl_1-\nl_2)=0$ which has solutions 
\BE
\nl_2=\nl_1\ \  \mbox{or}\ \ \nl_2=-3\nl_1.\label{a1sol}
\EE
Similarly, if $a_2\not= 0$,  $\cM_{211}\not=0$ then the ratio of equations (\ref{a2}) gives 
$(3\nl_2+\nl_1)(\nl_2-\nl_1)=0$ with solutions $\nl_2=\nl_1$ or $\nl_1=-3\nl_2$.
Thus for $a_1\not= 0$  and $a_2\not= 0$,  the only solution is $\nl_2=\nl_1=L/2$.

If only one $a_A\not= 0$, without loss of generality $a_1\not=0$ and $a_2=0$, then 
$\cM_{211}=0$ and from (\ref{tra}) $\cP_{211}=0$ and from (\ref{Bia5-f}) it follows
$\cP_{111}=0$. 
For the second solution in (\ref{a1sol}), $\nl_2=-3\nl_1$,  
 from (\ref{a1}) and (\ref{tra}) we obtain $a_1=-4\cM_{122}$ and $a_1=4\cM_{122}$, respectively, 
and thus this case does not occur.

If  $a_1=0=a_2$ then (\ref{a1}), (\ref{a2}) imply that all $\cM$'s  are zero and from (\ref{tra})
also all $\cP$'s  vanish, 
which leads to a type N or O spacetime.
\item
For $\Am=3$, the only components of $\Psi_{ijk}$, $\Psi_i$ and ${A_{ij}}$ are $\cM_{ABC}$, $a_A$ 
and $\cA_{AB}$, respectively.

From (\ref{CBS16}) with $C=B$ we obtain the equation 
\BE
a_A=\frac{2L+4\nl_A-4\nl_B}{\nl_B}\cM_{ABB}\mid_{B\not= A} 
\EE
that  together with (\ref{tra}) gives
\BEA
\fl
a_1&=&\frac{2L+4\nl_1-4\nl_2}{\nl_2}\cM_{122}=\frac{2L+4\nl_1-4\nl_3}{\nl_3}\cM_{133}=4(\cM_{122}+\cM_{133}),
\label{3m_a_1}\\
\fl
a_2&=&\frac{2L+4\nl_2-4\nl_1}{\nl_1}\cM_{211}=\frac{2L+4\nl_2-4\nl_3}{\nl_3}\cM_{233}=4(\cM_{211}+\cM_{233}),
\label{3m_a_2}\\
\fl
a_3&=&\frac{2L+4\nl_3-4\nl_1}{\nl_1}\cM_{311}=\frac{2L+4\nl_3-4\nl_2}{\nl_2}\cM_{322}=4(\cM_{311}+\cM_{322}).
\label{3m_a_3}
\EEA

(A) If all $a_A\not=0$ then the only solution is $\nl_1=\nl_2=L/2$ and $\nl_3=0$, which corresponds
to $\Am=2$ discussed above.

(B) Let us now study the case with one $a_A$ vanishing (we can assume $a_3=0$
without loss of generality). Then from (\ref{3m_a_3})
either $\cM_{311}+\cM_{322}=0$ and $\nl_1=\nl_2=L/2+\nl_3$, which implies $\Am=2$, $\nl_3=0$, $\nl_1=\nl_2=L/2$, or
$\cM_{311}=\cM_{322}=0$ and then
from (\ref{3m_a_1}) and (\ref{3m_a_2})  we get
\BEA
\ \ 3\nl_1^2-\nl_2^2-\nl_3^2+2(-\nl_1\nl_2 -\nl_1\nl_3+\nl_2\nl_3)&=&0, \label{eqfroma1}\\
-\nl_1^2+3\nl_2^2-\nl_3^2+2(-\nl_1\nl_2 +\nl_1\nl_3-\nl_2\nl_3)&=&0.
\EEA
Their difference gives
\BE
4(\nl_1-\nl_2)(\nl_1+\nl_2-\nl_3)=0
\EE
and thus either $\nl_1=\nl_2$ and from (\ref{eqfroma1})  $\nl_3=0$ or $\nl_3=\nl_1+\nl_2$
and  from (\ref{eqfroma1}) $\nl_1\nl_2=0$. This case thus again leads to $\Am=2$.

(C) If only one $a_A$ is non-vanishing (we again choose $a_1 \not=0$ and $a_2=a_3=0$ without loss of generality)
 then (\ref{3m_a_2}), (\ref{3m_a_3}) imply
either $\nl_1=L/2+\nl_3=\nl_2$ and $\nl_1=L/2+\nl_2=\nl_3$, which is not possible,
or $\nl_1=L/2+\nl_3=\nl_2$, which gives $\nl_1=\nl_2=L/2$ and $\nl_3=0$,
or $\nl_1=L/2+\nl_2=\nl_3$, which gives $\nl_1=\nl_3=L/2$ and $\nl_2=0$,
or $ \cM_{211}=\cM_{233}=\cM_{311}=\cM_{322}=0$. All these  cases, except the last one
are inconsistent with our assumption $\Am=3$. We thus need to check the last case
$\cM_{211}=\cM_{233}=\cM_{311}=\cM_{322}=0$, which has  two
possible branches corresponding to $\cM_{123}=0$ and $\cM_{123}\not=0$.

(a) $\cM_{123}=0$: from (\ref{CBS16}) (in the form (\ref{McabMcba}))
\BEA
(-\nl_1+3\nl_2+\nl_3)\cM_{123}&=&(-\nl_1+\nl_2+3\nl_3)\cM_{132},\label{MABC1}\\
\ \ (3\nl_1-\nl_2+\nl_3)\cM_{213}&=&(\nl_1-\nl_2+3\nl_3)\cM_{231},\label{MABC2}\\
\ \ (3\nl_1+\nl_2-\nl_3)\cM_{312}&=&(\nl_1+3\nl_2-3\nl_3)\cM_{321},\label{MABC3}
\EEA
we get $\nl_3=0$ and $\nl_1=L/2=\nl_2$  
for $\cM_{231}=\cM_{132}\not=0$,  
or $\cM_{231}=\cM_{132}=0$.
Then from (\ref{Bia3-a}) we get $\cA_{AB}=0$ and $a_1=\frac{4\nl_1}{\nl_2}\cM_{122}=\frac{4\nl_1}{\nl_3}\cM_{133}$
which together with (\ref{3m_a_1}) gives $\nl_2=\nl_3=L/2$, $\nl_1=0$. 

(b) $\cM_{123}\not=0$:
from (\ref{MABC1})--(\ref{MABC3})   and (\ref{Bia03}) we get 
\BE
\nl_1^2+\nl_2^2+\nl_3^2-2(\nl_1\nl_2+\nl_1\nl_3+\nl_2\nl_3)=0\label{MABC}
\EE
with the solution $\nl_3=\nl_1+\nl_2\pm 2\sqrt{\nl_1\nl_2}$. By substituting this result into (\ref{eqfroma1})
we get $\nl_1(\nl_2\pm\sqrt{\nl_1\nl_2})=0$ and thus either $\nl_1=0$ and $\nl_2=\nl_3=L/2$ or $\nl_1=\nl_2=L/2$
and $\nl_3=0 $ ($\nl_3=4\nl_1$ does not satisfy (\ref{eqfroma1})). 

(D) If all $a_A=0$ then from (\ref{3m_a_1})--(\ref{3m_a_3}) either one of $\nl_A=0$ and the other two are equal
to $L/2$ or all $\cM_{ABB}=0$. Then if 

a) $\cM_{123}=0$  from (\ref{MABC1})--(\ref{MABC3})   and (\ref{Bia03}) we get either  
one of $\nl_A=0$ and other two are again equal to $L/2$ or all $\cM_{ABC}=0$, which corresponds to a type N or O spacetime.

b) If $\cM_{123}\not=0$  then from (\ref{MABC1})--(\ref{MABC3})   and (\ref{Bia03}) we get (\ref{MABC}). 
Further from (\ref{Bia3-a})
we get all $\cA_{AB}=0$ and then 
\BEA
&&\nl_2\cM_{123}=\nl_3\cM_{132},\\
&&\nl_1\cM_{123}=-\nl_3\cM_{231},\\
&&\nl_1\cM_{132}=\nl_2\cM_{231},
\EEA
which yields
\BEA
&&0=(\nl_2-\nl_3)(\nl_1-\nl_2-\nl_3),\\
&&0=-\nl_1^2+\nl_1\nl_2+3\nl_1\nl_3+2\nl_2\nl_3-2\nl_3^2,\label{Bia3al2}\\
&&0=-\nl_1^2+3\nl_1\nl_2+\nl_1\nl_3+2\nl_2\nl_3-2\nl_2^2,
\EEA
with the solution either $\nl_1=\nl_2+\nl_3$ (then (\ref{Bia3al2}) implies $\nl_2$ or $\nl_3$ is zero)
or $\nl_2=\nl_3$ and (\ref{Bia3al2}) implies $\nl_1=0$ or $\nl_1=4\nl_2$; however, from
(\ref{Bia5-a}) we get $(\nl_2-\nl_1)\cM_{123}=0$ which is contradiction.

\end{enumerate}

\section{An example of a type D vacuum spacetime}
\label{typeD}

The five-dimensional rotating black-hole metric in Boyer-Lindquist coordinates
has the form \cite{Myers,Frolov} (we use the notation of \cite{Frolov})
\BEAH
\fl
\ \ \ \ \ {\rm d}s^2=&&\frac{\rho^2}{4 \Delta} {\rm d}x^2 + \rho^2 {\rm d} \theta^2 -{\rm d}t^2 + (x+a^2) \sin^2 \theta
{\rm d} \phi^2 + (x+b^2) \cos^2 \theta {\rm d} \psi^2  \\
\fl
&& + \frac{{r_0}^2}{\rho^2} ({\rm d}t + a \sin^2 \theta {\rm d} \phi
+ b \cos^2 \theta {\rm d} \psi)^2 ,
\EEAH
where
\BDM
\rho^2=x+a^2 \cos^2 \theta + b^2 \sin^2 \theta ,\ \ 
\Delta=(x+a^2)(x+b^2)-{r_0}^2 x .
\EDM
After normalizing two null vectors $L_+$ and $L_-$ given in (4.23) in \cite{Frolov}, one
can appropriately choose the rest of the frame vectors $\bm^1$, $\bm^2$ and $\bm^3$.
It turns out that all components of the Weyl tensor with the boost weights 2, 1, -1, -2 vanish
and hence the spacetime is  of the algebraic type D \cite{Algclass,Weylletter}.
One can also explicitly calculate the matrix $\mS$. While the form of $\mS$ depends on the choice of
$\bm^1$, $\bm^2$ and $\bm^3$, the characteristic polynomial of $\mS$, $P_{\lambda}(\mS)$, does not.
For $P_{\lambda}(\mS)$ we obtain
\BE
P_{\lambda}(\mS) = \left(\lambda-\frac{1}{\sqrt{x}}\right) \left(\lambda-\frac{\sqrt{x}}{\rho^2}  \right)^2
\EE
and the diagonal form of $\mS$ is
\BE
\mS_{\rm{diag}}= \left( \begin {array}{ccc} {\frac {1}{\sqrt {x}}}&0&0
\\\noalign{\medskip}0&{\frac {\sqrt {x}}{{\rho}^{2}}}&0
\\\noalign{\medskip}0&0&{\frac {\sqrt {x}}{{\rho}^{2}}}\end {array}
 \right) 
\EE  
From (\ref{shearmatrix}) it follows that shear is
\BE
\sigma = \sqrt{\frac{2}{3x}} \left(1-\frac{x}{\rho^2}\right),
\EE
which is in accordance with equation (4.25) in \cite{Frolov}. Note that for $a=b=0$ all eigenvalues
of $\mS$ are equal and  shear is thus zero. Whereas for $a^2+b^2>0$
the matrix $\mS$ has two equal 
and one distinct eigenvalue and shear does not vanish. Note also that similarly as for the types III and N,
this algebraically special vacuum solution has non-vanishing shear (see also footnote 4 in paper \cite{Frolov}).
Moreover, in this spacetime $\mS$ does not have properties
proved in Sections \ref{typeN} and \ref{PIII} for vacuum type N and type III spacetimes.

\end{document}